\documentclass[1p]{elsarticle}

\usepackage{lineno,hyperref,amsmath}
\usepackage{url}
\usepackage{graphicx}
\usepackage{epstopdf,amsmath,color,appendix}
\modulolinenumbers[5]

\journal{Journal of \LaTeX\ Templates}

\begin{document}

\begin{frontmatter}

\title{Connectionist-Symbolic Machine Intelligence using Cellular Automata based Reservoir-Hyperdimensional Computing}

\author[mymainaddress]{Ozgur Yilmaz\corref{mycorrespondingauthor}}
\cortext[mycorrespondingauthor]{Corresponding Author}
\ead{ozyilmaz@turgutozal.edu.tr, Website: ozguryilmazresearch.org}

\address[mymainaddress]{Turgut Ozal University, Department of Computer Engineering, Ankara Turkey}

\begin{abstract}
We introduce a novel framework of reservoir computing, that is capable of both connectionist machine intelligence and symbolic computation. Cellular automaton is used as the reservoir of dynamical systems. Input is randomly projected onto the initial conditions of automaton cells and nonlinear computation is performed on the input via application of a rule in the automaton for a period of time. The evolution of the automaton creates a space-time volume of the automaton state space, and it is used as the reservoir. The proposed framework is capable of long short-term memory and it requires orders of magnitude less computation compared to Echo State Networks. We prove that cellular automaton reservoir holds a distributed representation of attribute statistics, which provides a more effective computation than local representation. It is possible to estimate the kernel for linear cellular automata via metric learning, that enables a much more efficient distance computation in support vector machine framework. Also, binary reservoir feature vectors can be combined using Boolean operations as in hyperdimensional computing, paving a direct way for concept building and symbolic processing.
\end{abstract}


\end{frontmatter}

\section{Introduction}
\begin{figure}
\begin{center}
\includegraphics[width=0.5\textwidth]{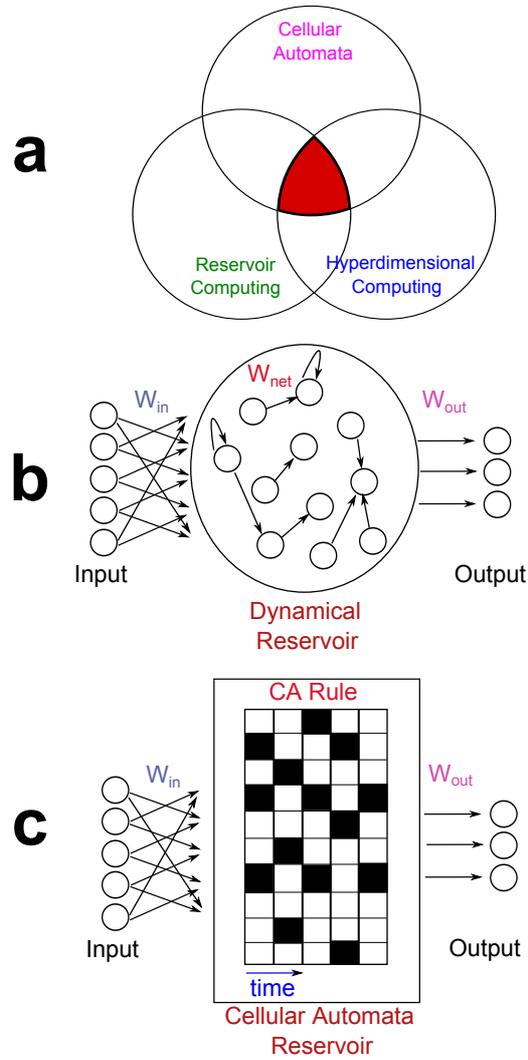}
\end{center}
   \caption{\textbf{a.} Our work is a cross fertilization of cellular automata, reservoir computing and hyperdimensional computing frameworks. \textbf{b.} In classical reservoir computing, data is projected on a randomly generated recurrent neural network and the activities of the network are harvested which is called reservoir. \textbf{c.} In cellular automata reservoir, data is projected onto cellular automaton instead of a neural network. }
\label{fig:figm1}
\end{figure}

We introduce a novel framework of cellular automata based reservoir computing, that is capable of long short-term memory. Cellular automaton is used as the reservoir of dynamical systems. Input is randomly projected onto the initial conditions of automaton cells and nonlinear computation is performed on the input via application of a rule in the automaton for a period of time. The evolution of the automaton creates a space-time volume of the automaton state space, and it is used as the feature vector. The proposed framework requires orders of magnitude less computation compared to Echo State Networks. We prove that cellular automaton reservoir holds a distributed representation of attribute statistics, which provides a more effective computation than local representation. It is possible to estimate the kernel for linear cellular automata via metric learning, that enables a much more efficient distance computation in support vector machines framework. Additionally, the binary nature of the feature space and additivity of the cellular automaton rules enable Boolean logic, and provides a potential for symbolic processing. Our study is a cross fertilization of cellular automata, reservoir computing and hyperdimensional computing frameworks (Figure \ref{fig:figm1} a). 

In the following sections we review reservoir computing, cellular automata and neuro-symbolic computation, then we state the contribution of our study.  

\subsection{Reservoir Computing}
Recurrent Neural Networks are connectionist computational models that utilize distributed representation with nonlinear units. Information in RNNs is propagated and processed in time through the states of its hidden units, which make them appropriate tools for sequential information processing. 

RNNs are known to be Turing complete computational tools \cite{siegelmann1995computational} and universal approximators of dynamical systems \cite{funahashi1993approximation}. They are especially appealing for problems that require remembering long-range statistical relationships such as speech, natural language processing, video processing, financial data analysis etc. 

Despite their immense potential as universal computers, problems arise in training RNNs due to the inherent difficulty of learning long-term dependencies \cite{hochreiter1991untersuchungen,bengio1994learning,hochreiter1997long} and convergence issues \cite{doya1992bifurcations}. However, recent advances suggest promising approaches in overcoming these issues, such as emulating the recurrent computation in neural network framework (\cite{yilmazFastRecurrent} under review) or utilizing a reservoir of coupled oscillators \cite{jaeger2001echo,maass2002real}.

Reservoir computing (a.k.a. echo state networks or liquid state machines) alleviates the problem of training a recurrent network by using a randomly connected dynamical reservoir of coupled oscillators, which are operating at the edge of chaos. It is claimed that many of these type of dynamical systems possess high computational power \cite{bertschinger2004real,legenstein2007edge}. In this approach, due to rich dynamics already provided by the reservoir, there is no need to train many recurrent layers and learning takes place only at the output (read-out) layer. This simplification enables usage of recurrent neural networks in complicated tasks that require memory for long-range (both spatially and temporally) statistical relationships. It can be considered in a larger family of architectures where random connections are utilized instead of trained ones, another popular feedorward algorithm is named extreme learning machines \cite{huang2006extreme}. \footnote{The comparison of the two approaches are given in \cite{butcher2013reservoir}.}

The essential feature for stability of the randomly connected network is called echo state property \cite{jaeger2001echo}. In networks with this property, the effect of previous state and previous input dissipates gradually in the network without getting amplified. In classical echo state networks, the network is generated randomly and sparsely, considering the spectral radius requirements of the weight matrix. Even though spectral radius constraint ensures stability of the network to some extent, it does not say anything about the short-term memory or computational capacity of the network. The knowledge about this capacity is essential for proper design of the reservoir for the given task. 

The reservoir is expected to operate at the edge of chaos because the dynamical systems are shown to present high computational power at this mode \cite{bertschinger2004real,legenstein2007edge}. High memory capacity is also shown for reservoirs at the edge of chaos. Lyapunov exponent is a measure edge of chaos operation in a dynamical system, and it can be empirically computed for a reservoir network \cite{legenstein2007edge}.  However, this computation is not trivial or automatic, and needs expert intervention \cite{lukovsevivcius2009reservoir}. 

It is empirically shown that there is an optimum Lyapunov exponent of the reservoir network, related to the amount of memory needed for the task \cite{verstraeten2007experimental}. Thus, fine-tuning the connections in the reservoir for learning the optimal connections that lead to optimal Lyapunov exponent is very crucial for achieving good performance with the reservoir. There are many types of learning methods proposed for tuning the reservoir connections (see \cite{lukovsevivcius2009reservoir} for a review), however optimization procedure on the weight matrix is prone to get stuck at local optimum due to high curvature in the weight space. 
 
The input in a complex task is generated by multiple different processes, for which the dynamics and spatio-temporal correlations might be very different. One important shortcoming of the classical reservoir computing approach is its inability to deal with multiple spatio-temporal scales simultaneously. Modular reservoirs have been proposed that contain many decoupled sub-reservoirs operating in different scales, however fine tuning the sub-reservoirs according to the task is a non-trivial task. 

\begin{figure*}
\begin{center}
\includegraphics[width=1\textwidth]{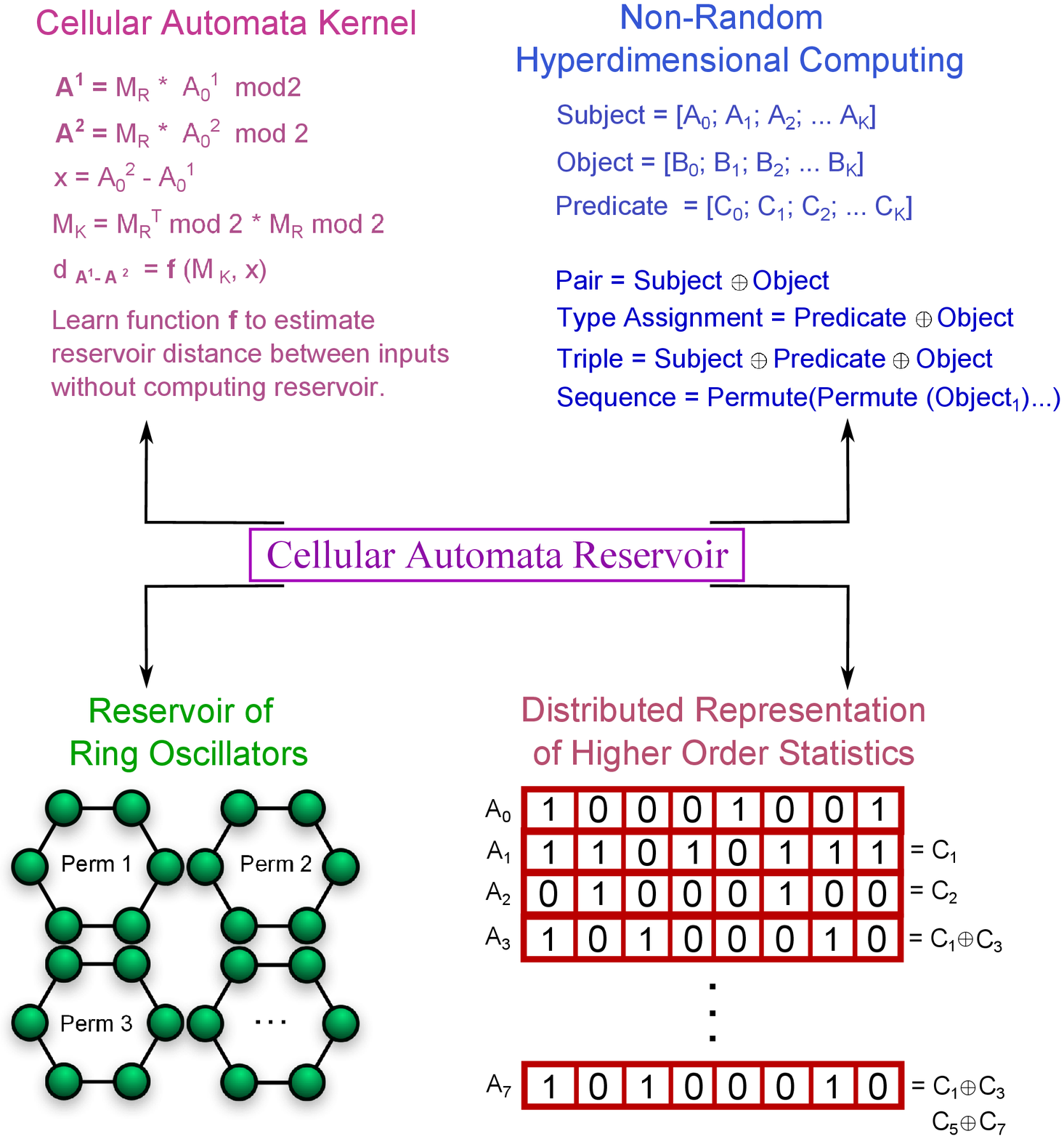}
\end{center}
   \caption{Four aspects of cellular automata based reservoir computing.  \textbf{Lower Left.} Cellular automata reservoir is essentially an ensemble of ring oscillators because of one-neighbor connectivity of cellular automata.  \textbf{Lower Right.} The cellular automata reservoir holds a distributed representation of high order attribute statistics (section \ref{sec:distributedRep}), that is shown to be superior over local representation. $A_0$ is the initial condition for CA and it evolves in time. Whole space time evolution, $A_k$, is the reservoir, and it holds higher order attribute statistics given as $C_k$. \textbf{Upper Left.} The computation in linear cellular automaton rules can be kernelized to be used in SVM framework (section \ref{sec:kernellearning_section}), where the kernel performance approaches nonlinear kernel performance with only linear computational complexity.  \textbf{Upper Right.} High dimensional binary vectors of CA reservoir can be used for symbolic processing, as in hyperdimensional computing, where we can use predicates, define triples, sequences and composite objects. This capability allows simulataneous connectionist and symbolic computation in our framework.} 
\label{fig:fig0}
\end{figure*}

\subsection{Cellular Automata}
Cellular automaton is a discrete computational model consisting of a regular grid of cells, each in one of a finite number of states (Figure \ref{fig:figm1} c). The state of an individual cell evolves in time according to a fixed rule, depending on the current state and the state of neighbors. The information presented as the initial states of a grid of cells is processed in the state transitions of cellular automaton and computation is typically very local. Some of the cellular automata rules are proven to be computationally universal, capable of simulating a Turing machine \cite{cook2004universality}. 

The rules of cellular automata are classified \cite{wolfram2002new} according to their behavior: attractor, oscillating, chaotic, and edge of chaos. Turing complete rules are generally associated with the last class (rule 110, Conway’s game of life). Lyapunov exponent of a cellular automaton can be computed and it is shown to be a good indicator of computational power of the automata \cite{baetens2010phenomenological}. A spectrum of Lyapunov exponent values can be achieved using different cellular automata rules. Therefore a dynamical system with specific memory capacity (i.e. Lyapunov exponent value) can be constructed by using a corresponding cellular automaton (or a hybrid automaton \cite{sipper1998evolving}).  

Cellular automata have been previously used for associative memory and classification tasks. Tzionas et al. \cite{tzionas1994new} proposed a cellular automaton based classification algorithm. Their algorithm clusters 2D data using cellular automata, creating boundaries between different seeds in the 2D lattice. The partitioned 2D space creates geometrical structure resembling a Voronoi diagram. Different data points belonging to the same class fall into the same island in the Voronoi structure, hence are attracted to the same basin. Clustering property of cellular automata is exploited in a family of approaches, using rules that form attractors in lattice space \cite{chady1998evolution,ganguly2003survey}. The attractor dynamics of cellular automata resembles Hopfield network architectures \cite{hopfield1982neural}. The time evolution of the cellular automata has very rich computational representation, especially for edge of chaos dynamics, but this is not exploited if the presented data are classified according to the converged basin in 2D space. To alleviate this shortcoming, the proposed algorithm in this paper exploits the \textbf{entire} time evolution of the CA, and uses the states as the reservoir of nonlinear computation.  

Another cellular based machine learning approach is cellular neural networks \cite{chua1988cellular}. It has been shown that every binary cellular automata of any dimension is a special case of a cellular neural network of the same neighborhood size \cite{chua2002nonlinear}. However, cellular neural networks impose a very specific spatial structure and they are generally implemented on specialized hardware, generally for image processing (see \cite{javier2013efficient} for a recent design).

A relatively new class of cellular automata approach is Elementary Cellular Automata with Memory (ECAM) \cite{alonso2006elementary,martinez2013designing}. The state of each cell is determined by the history of the cell using majority, minority or parity (XOR) rules. This line of work is very related with the proposed framework in this paper (section \ref{sec:distributedRep} on distribted representation of statistics) and a cross-talk between reservoir computing and ECAM seems essential.

\subsection{Symbolic Computation on Neural Representations}
Uniting the expressive power of mathematical logic and pattern recognition capability of neural networks has been an open question for decades, although several successful theories have been proposed \cite{laird1987soar,anderson2014atomic,pollack1990recursive,van2006neural,bader2008connectionist}. The grand challenge is systematically tackled by a large group of scientists, and the designs are most of the time inspired by cortical models \cite{samsonovich2012roadmap}. There are many theories, but a dominant solution is not proposed yet. Difficulty arises due to the very different mathematical nature of logical reasoning and dynamical systems theory. We conjecture that combining connectionist and symbolic processing requires \textbf{commonalizing the representation of data and knowledge}. Recently Jaeger proposed a novel framework called "Conceptors" based on reservoir computing architecture \cite{jaeger2014controlling}. The Conceptors are linear operators learned from the activities of the reservoir neurons and they can be combined by elementary logical operators, which enables them to form symbolic representations of the neural activities and build semantic hierarchies. In a similar flavor, Mikolov et al. \cite{mikolov2013distributed} successfully used neural network representations of words (language modeling) for analogical reasoning. 

Kanerva introduced hyperdimensional computing \cite{kanerva2009hyperdimensional} that utilizes high dimensional random binary vectors for representing objects and predicates for symbolic manipulation and inference. \footnote{The general family of the approach is called 'reduced representations', and detailed introduction can be found in \cite{plate2003holographic}} In this approach, high dimensionality and randomness enable binding and grouping operations (see \cite{snaider2012integer,snaider2012extended} for extensions.) that are essential for one shot learning, analogy making and hierarchical concept building. Most recently Galant et al. \cite{gallant2013representing} introduced random matrices to this context and extended the binding and quoting operations. The two basic mathematical tools of reduced representations are vector addition and XOR. In this paper we borrow these tools of hyperdimensional computing framework, and build a semantically more meaningful representation by removing the randomness and replacing it with cellular automata computation. The two main mathematical tools are vector addition and XOR. In this paper we borrow the tools of hyperdimensional computing framework, and build a semantically more meaningful representation by removing the randomness and replacing it with cellular automata computation.

\subsection{Contributions}
We first provide a feedorward architecture using cellular automata computation. Then we show that the proposed architecture is capable of accomplishing long-short-term-memory tasks such as the 5 bit and 20 bit noiseless memorization  (Table \ref{table:5BitElem}), which are known to be problematic for feedforward architectures \cite{hochreiter1997long}. We prove that the computation performed by cellular automata produces a distributed representation of higher order attribute statistics, which gives dramatically superior performance over local representation (Table \ref{table:5BitElemCA_LocalRep} and \ref{table:MNIST_KernelTwoFeatures}). Also we demonstrate that cellular automata feature expansion for linear rules can be kernelized via metric learning, and the estimated kernel significantly reduces computation in both training and test stages in support vector machines while the performance approaches nonlinear methods (Table \ref{table:MNIST_KernelTwoFeatures}).

Then we provide a low computational complexity method (Table \ref{table:ComparisonOperation}) for implementing reservoir computing based recurrent computation, using cellular automata. Cellular automata replace the echo state neural networks. This approach provides both theoretical and practical advantages over classical neuron-based reservoir computing.  The computational complexity of the feedforward and recurrent framework is orders of magnitude lower than echo state network based reservoir computing approaches (Table \ref{table:ComparisonOperation}). 

The classification performance of the cellular automata (CA) feature is also examined on CIFAR 10 dataset \cite{krizhevsky2009learning} in section \ref{sec:BinarizeData}, in which we show that CA features exceed the performance of RBF kernel with fraction of the computation. The computational complexity of the framework is orders of magnitude lower than echo state network based reservoir computing approaches (Table \ref{table:ComparisonOperation} ). 

Additionally we show that the framework has potential for symbolic processing such that the cellular automata feature space can directly be combined by Boolean operations as in hyperdimensional computing, hence they can represent concepts and form a hierarchy of semantic interpretations. We demonstrate this capability by making analogies directly on images (Tables \ref{table:AnalogyCA} and \ref{table:AnalogyCA2}). The versatility of our framework is illustrated in Figure \ref{fig:fig0}. In the next section we give the details of the algorithm and then provide results on several simulations and experiments that demonstrate our contributions. \footnote{The details of each the algorithms and experiments are given separately in an appendix.}

\section{Algorithm}
\label{sec:method_section}
In our system, data are passed on a cellular automaton instead of an echo state network and the nonlinear dynamics of cellular automaton provide the necessary projection of the input data onto an expressive and discriminative space.  Compared to classical neuron-based feedforward or reservoir computing, the feature space design is trivial: cellular automaton rule selection. Utilization of ‘edge of chaos’ automaton rules ensures Turing complete computation in the feature space, which is not guaranteed in classical neural network approaches. 

\textbf{As a starting point we first provide a feedforward architecture.} The analysis of the feedforward architecture gives a valuable intuition for the capabilities of the cellular automaton computation. For this purpose we look into memorization capability of the feedforward algorithm in section \ref{sec:results_section}, then analyze the distributedness of the representation in section \ref{sec:distributedRep}. We kernelize the feedforward feature space in sections \ref{sec:metriclearning_section} and \ref{sec:kernellearning_section}. The neuralization of the feedforward architecture is introduced in section \ref{sec:NetworkArchitecture}. The recurrent architecture is given in section \ref{sec:CA_Reservoir_Recurrent}, and until that section all the arguments in the paper are made for the feedforward cellular automata feature expansion. The added capability of due to recurrence is analyzed in detail, and experimental results are presented in section \ref{sec:CA_Reservoir_Recurrent}.   

\begin{figure*}
\begin{center}
\includegraphics[width=1\textwidth]{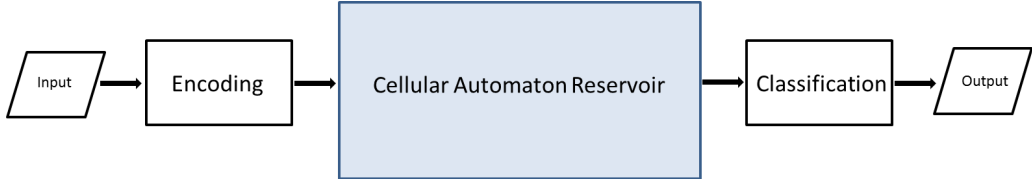}
\end{center}
   \caption{General framework for cellular automata based computing.}
\label{fig:fig1}
\end{figure*}

Algorithmic flow is shown in Figure \ref{fig:fig1}. The encoding stage translates the input into the initial states of a 1D or multidimensional cellular automaton (2D is shown as an example). In cellular automaton computing stage, the cellular automaton rules are executed for a fixed period of iterations ($I$), to evolve the initial states. The evolution of the cellular automaton is recorded such that, at each time step a snapshot of the whole states in the cellular automaton is vectorized and concatenated. This output is a projection of the input onto a nonlinear cellular automata state space. Then the  cellular automaton output is used for further processing according to the task (eg. classification, compression, clustering etc.).

\begin{figure}
\begin{center}
\fbox{\includegraphics[width=1\textwidth]{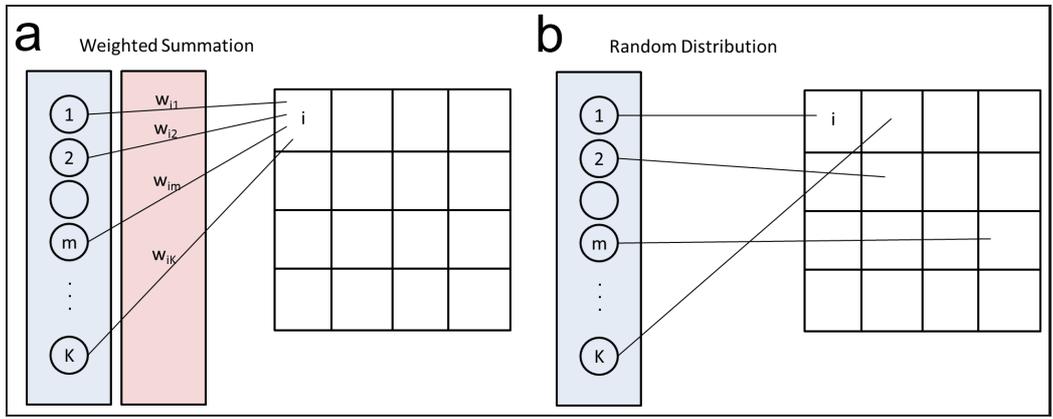}} 
\end{center}
   \caption{Two options for encoding the input into cellular automaton initial states. \textbf{a}. Each cell receives a weighted sum of the input dimensions. \textbf{b}. Each feature dimension is randomly mapped onto the cellular automaton cells. This last option is adopted for all the experiments in the paper. }
\label{fig:fig2}
\end{figure}

In encoding stage there are two proposed options depending on the input data. 
\textbf{1.} For non-binary input data, each cell of cellular automaton might receive weighted input from every feature dimension of the input (Figure \ref{fig:fig2}a). The weighted sum is then binarized for each cell. In this option, instead of receiving input from the whole set of feature dimensions, a single cell can receive input from a subset of feature dimensions. In that case, the weight vector for a cell is sparse and a subspace of the input is processed by specific cells. In general, the weights can be set randomly as in echo state networks. 
\textbf{2.} For binary input data, each feature dimension can randomly be mapped onto the cells of the cellular automaton (Figure \ref{fig:fig2}b). The size of the CA should follow the input feature dimension. For the sake of simplicity we adopted the second option throughout the paper. 

\begin{figure}
\begin{center}
\includegraphics[width=1\textwidth]{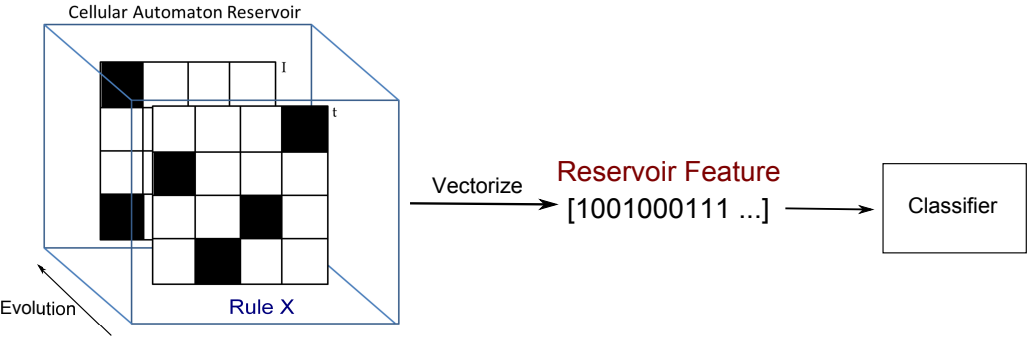} 
\end{center}
   \caption{Cellular automaton (CA) feature space which is the space-time volume of the automaton evolution using Rule X. The whole evolution of the CA is vectorized and it is used as the feature vector for classification.}
\label{fig:fig3}
\end{figure}

After encoding with second option, suppose that cellular automaton is initialized with vector ${A_0}^{P_1}$, in which $P_1$ corresponds to a permutation of raw input data. Then cellular automata evolution is computed using a prespecified rule, $Z$ (Figure \ref{fig:fig3}):
\begin{flalign*}
    &{A_1}^{P_1} = Z({A_0}^{P_1}), \\
		&{A_2}^{P_1} = Z({A_1}^{P_1}), \\
		&...\\
		&{A_I}^{P_1} = Z({A_{I-1}}^{P_1}).
\end{flalign*}  
We concatenate the evolution of cellular automata to obtain a vector for a single permutation:
\begin{flalign*}
		&{A}^{P_1} = [{A_0}^{P_1}; {A_1}^{P_1}; {A_2}^{P_1}; ... {A_I}^{P_1}]
\end{flalign*}

It is experimentally observed that multiple random mappings significantly improve accuracy. There are $R$ number of different random mappings, i.e. separate CA feature spaces, and they are combined into a large feature vector:
\begin{flalign*}
		&\mathbf{{A}^{R}} = [{A}^{P_1}; {A}^{P_2}; {A}^{P_3}; ... {A}^{P_R}].
\end{flalign*} 

The computation in CA takes place when cell activity due to nonzero initial values (i.e. input) mix and interact. Both prolonged evolution duration and existence of different random mappings increase the probability of long range interactions, hence improve computational power. We analyze the effect of number of evolutions and number of random mappings on computational capacity in Section \ref{sec:distributedRep}. 

\section{Memory of Cellular Automata State Space}
\label{sec:results_section}
In order to test for long-short-term-memory capability of the proposed feedforward cellular automata framework, 5 bit and 20 bit memory tasks were used. They are proposed to be problematic for feedforward architectures, also reported to be the hardest tasks for echo state networks in \cite{jaeger2012long}. In these tasks, a sequence of binary vectors are presented, then following a distractor period, a cue signal is given after which the output should be the initially presented binary vectors. Input-output mapping is learned by estimating the linear regression weights via pseudo-inverse (see \cite{jaeger2012long} for details). These tasks have binary input, hence it is possible to randomly map the input values onto cells as initial states (Figure \ref{fig:fig2} b). In echo state networks, a sequence is presented to the network one step at a time, while the network is remembering the previous inputs in its holistic state. In our feedforward architecture, sequence is flattened: all the sequence data are presented to the network at once (vectorization) and the whole output sequence is estimated using regression. Increasing the distractor period expands the size of the vector, and adds irrelevant data to the CA computation reducing the estimation quality. 

Using a larger data portion for context is common in sequence learning and it is called sliding window method, and estimating the whole output sequence is generally utilized in hidden Markov model approaches \cite{dietterich2002machine}, but these two were not applied together before to the best of our knowledge. Estimating the whole sequence using the whole input sequence is a much harder task, and we are making our case more difficult in order to keep the comparison with literature (explained in more detail in section \ref{sec:CA_Reservoir_Recurrent}). 

Both 1D elementary CA rules and 2D Game of Life CA is explored. The total computational complexity is determined by the number of different random mappings, $R$, (i.e. separate feature spaces, Figure \ref{fig:fig3}) and the number of cell iterations $I$. The success criteria and regression method provided by \cite{jaeger2012long} are used in the experiments.   

\subsection{Game of Life}
5 bit task is run for distractor period $T_0$ 200 and 1000. The percent of trials that failed is shown for various $R$ and $I$ combinations in Table \ref{table:5BitGoL200and1000}. 20 bit task is run for distractor period $T_0$ 200 and 300, and again percent failed trials is shown in Table \ref{table:20BitGoL200and300}. It is observed that Game of Life CA is able to solve both 5 bit and 20 bit problems and for very long distractor periods. 
              
\begin{table}[h]                 
\centering                        
\begin{tabular}{l|l|l|l|l} 
$T_0 = 200$ & R=4 & 16 & 32 & 64 \\  
\hline     
I=4 & 100 & 100 & 100 & 4 \\
16 & 100 & 28 & 0 & 0 \\  
32 & 100 & 0 & 0 & 0 \\   
64 & 100 & 0 & 0 & 0 \\   
\end{tabular}                            
\quad                            						                                   
\begin{tabular}{l|l|l|l|l}   
 $T_0 = 1000$ & R=4 & 16 & 32 & 64 \\  
\hline      
I=4 & 100 & 100 & 100 & 100 \\
16 & 100 & 100 & 100 & 14 \\
32 & 100 & 100 & 100 & 0 \\ 
64 & 100 & 100 & 22 & 0 \\  
\end{tabular}                     
\caption{Percent failed trials for 5 Bit Task, Game of Life CA, $T_0 = 200$ (Left) $T_0 = 1000$ (Right). Rows are number of iterations $I$, and Columns are number of random permutations.}          
\label{table:5BitGoL200and1000}        
\end{table} 

\begin{table}[h]           
\centering                  
\begin{tabular}{l|l|l|l|l}      
 $T_0 = 200$ & R=192 & 256 & 320 & 384 \\ 
\hline  
I=12 & 100 & 100 & 100 & 32 \\
16 & 100 & 84 & 12 & 0 \\   
\end{tabular}                
\quad  
\begin{tabular}{l|l|l|l|l}      
 $T_0 = 300$ & R=192 & 256 & 320 & 384 \\ 
\hline  
I=12 & 100 & 100 & 100 & 60 \\
16 & 100 & 100 & 20 & 0 \\  
\end{tabular}               
\caption{Percent failed trials for 20 Bit Task, Game of Life CA, $T_0 = 200$ (Left) $T_0 = 300$ (Right). Rows and columns are the same as above.}    
\label{table:20BitGoL200and300}  
\end{table} 

\subsection{Elementary Cellular Automata}
5 bit task ($T_0 = 200$) is used to explore the capabilities of elementary cellular automata rules. Rules 32, 160, 4, 108,  218 and 250 are unable to give meaningful results for any [$R$, $I$] combination. Rules 22, 30, 126, 150, 182, 110, 54, 62, 90, 60 are able give 0 error for some combination. Best performances are observed for rules  90, 150, 182 and 22, in decreasing order (Table \ref{table:5BitElem}). It is again observed that computational power is enhanced with increasing either $R$ or $I$, thus multiplication of the two variables determine the overall performance. 

\begin{table}[h]           
\centering                  
\begin{tabular}{l|l|l|l|l}      
 $Rule 90$ & R=8 & 16 & 32 & 64 \\ 
\hline 
I=8 & 100 & 78 & 12 & 0 \\  
16 & 74 & 4 & 0 & 0 \\    
32 & 4 & 2 & 0 & - \\     
\end{tabular}                
\quad  
\begin{tabular}{l|l|l|l|l}      
 $Rule 150$ & R=8 & 16 & 32 & 64 \\ 
\hline 
I=8 & 100 & 80 & 8 & 0 \\   
16 & 84 & 6 & 0 & 0 \\    
32 & 8 & 0 & 0 & - \\
\end{tabular}                
\quad  
\begin{tabular}{l|l|l|l|l}      
 $Rule 182$ & R=8 & 16 & 32 & 64 \\ 
\hline 
I=8 & 100 & 82 & 18 & 0 \\  
16 & 92 & 14 & 0 & 0 \\   
32 & 12 & 0 & 0 & - \\  
\end{tabular}  
\quad
\begin{tabular}{l|l|l|l|l}      
 $Rule 22$ & R=8 & 16 & 32 & 64 \\ 
\hline 
I=8 & 100 & 78 & 20 & 0 \\  
16 & 86 & 16 & 0 & 0 \\   
32 & 16 & 0 & 0 & - \\   
\end{tabular} 
\caption{Percent failed trials for 5 Bit Task, Elementary CA Rules, $T_0 = 200$. Rows and columns are the same as above. }    
\label{table:5BitElem}  
\end{table} 

Previous studies have shown that, even recurrent architectures are having hard time to accomplish these tasks, while cellular automata feature expansion is able to give zero error. Thus, the cellular automaton state space seems to offer a rich feature space, that is capable of long-short-term-memory. Interestingly, class 3 CA rules \cite{wolfram2002new} which show random behavior seem to give best performance in this memorization task. \footnote{The results and arguments in the following sections will always be derived using elementary cellular automata. }

\section{Distributed Representation of Higher Order Statistics}
\label{sec:distributedRep}
Discovering the latent structure in high dimensional data is at the core of machine learning, and capturing the statistical relationship between the data attributes is of the essence \footnote{Sometimes this is called learning 'long range dependencies'.}. Second order statistics (covariance matrix) is shown to be very informative for a set of AI tasks \cite{tuzel2006region}. However, higher order statistics are known to be required for a wide variety of problems \cite{mendel1991tutorial,aggarwal2000finding}. Polynomial kernels in support vector machine framework represent these statistics in a local manner \cite{goldberg2008splitsvm}. The computation in cellular automata can be viewed from many perspectives using a variety of mathematical formalisms \cite{mitchell1996computation}. We are approaching this from a machine learning path, and trying the understand the meaning of cellular automaton evolution in terms of attribute statistics \footnote{To calirfy the language: the initial vector $A_0$ of CA hold the data attributes, and the evolution of CA results in CA features.}. We show that linear cellular automata compute higher order statistics and hold them within its states in a distributed manner, which shows clear advantages over local representation (see Table \ref{table:5BitElemCA_LocalRep} and \ref{table:MNIST_KernelTwoFeatures} ). 

For simplification and due to the encouraging results from the aforementioned memory experiments (Table \ref{table:5BitElem}), we analyzed a linear automaton, rule 90. Suppose that cellular automaton is initialized with vector $A_0$, which holds the attributes of the raw data. The cellular automaton activity at time $1$ is given by:
\begin{flalign*}
    &{A}_{1} = {\Pi}_{1} {A}_{0} \oplus {\Pi}_{-1} {A}_{0}, \\
\end{flalign*} 
where ${\Pi}_{1}$ and ${\Pi}_{-1}$ are matrices $+1$ and $-1$ bit shift operations and $\oplus$ is bitwise XOR.\\
Similarly,
\begin{flalign*}
    &{A}_{2} = {\Pi}_{2} {A}_{0} \oplus {\Pi}_{-2} {A}_{0}, \\
		&{A}_{3} = {\Pi}_{3} {A}_{0} \oplus {\Pi}_{-3} {A}_{0} \oplus {\Pi}_{1} {A}_{0} \oplus {\Pi}_{-1} {A}_{0}, \\
		&{A}_{4} = {\Pi}_{4} {A}_{0} \oplus {\Pi}_{-4} {A}_{0}, \\
		&{A}_{5} = {\Pi}_{5} {A}_{0} \oplus {\Pi}_{-5} {A}_{0} \oplus {\Pi}_{3} {A}_{0} \oplus {\Pi}_{-3} {A}_{0}, \\
		\text{. . .}
\end{flalign*} 
See Figure \ref{fig:fig0} lower right, for visualizing the evolution.

\begin{figure}
\begin{center}
\includegraphics[width=0.5\textwidth]{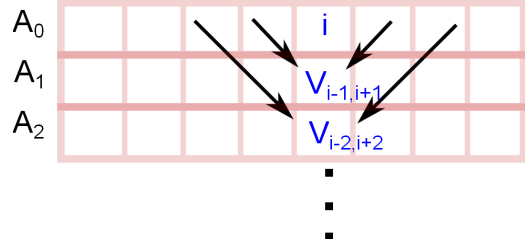} 
\end{center}
   \caption{The cell value for rule 90 (i.e. $V_{i-k,i+k}$) represents covariance of initial conditions (attributes), that are $2k$ apart from each other, where $k$ is the time step of evolution. This is demonstrated for a single column of the state evolution. }
\label{fig:fig3p5}
\end{figure}

It can be shown that by induction:
\begin{flalign*}
    &{A}_{k} = {\Pi}_{k} {A}_{0} \oplus {\Pi}_{-k} {A}_{0} \oplus LOT,  \\
\end{flalign*} 
where $LOT$ are the lower order terms, i.e. ${\Pi}_{i} {A}_{0} \oplus {\Pi}_{-i} {A}_{0}, \text{ for } i<k$. \\
Let us define:
\begin{flalign*}
    &{C}_{k} = {\Pi}_{k} {A}_{0} \oplus {\Pi}_{-k} {A}_{0},  \\
\end{flalign*} 
which is nothing but the covariance of the attributes that are $2k$ apart from each other, which is hold at the center bit of $k^{th}$ time step of cellular automaton evolution (Figure \ref{fig:fig3p5}). 

Then:
\begin{flalign*}
    &{A}_{1} = {C}_{1},  \\
		&{A}_{2} = {C}_{2}, \\
		&{A}_{3} = {C}_{3} \oplus {A}_{1}, \\
		&{A}_{4} = {C}_{4}, \\
		&{A}_{5} = {C}_{5} \oplus {A}_{3} \oplus {A}_{1} \\
		&\text{. . .}
\end{flalign*} 
Using induction:
\begin{flalign*}
    &{C}_{k} = {A}_{k} \oplus \prod_{i \in S}^{} A_{i},  \\
\end{flalign*} 
where the set $S$ holds lower order terms (i.e. cellular automaton time steps before $k$) and product symbol represents consecutive XOR operations over the defined set of cellular automaton states.\\

Therefore we can deduce any attribute covariance $C_k$, via proper application of XOR on cellular automaton state evolution, $A_k$. Otherwise, the cellular automaton states hold a \textbf{distributed representation of attribute higher order statistics}:
\begin{flalign*}
    &{A}_{k} = {C}_{k} \oplus \prod_{i \in S}^{} C_{i}.  \\
\end{flalign*} 

What is the function of multiple initial random permutations? Initial random permutation (eg. $R_1$) used in our framework is an extra matrix multiplication on initial vector $A_0$, then time evolution equation is:
\begin{flalign*}
    &{A}_{1} = {\Pi}_{1} {\Pi}^{R_1} {A}_{0} \oplus {\Pi}_{-1} {\Pi}^{R_1} {A}_{0}, \\
\end{flalign*} 
and the two matrices can be combined into one:
\begin{flalign*}
    &{\Pi}_{1}^{R_1} = {\Pi}_{1} {\Pi}^{R_1}. \\
\end{flalign*} 
Without random permutation, the $i^{th}$ bit at time step $1$ holds the covariance between ${i-1}^{st}$ and ${i+1}^{st}$ attributes. This is illustrated in Figure \ref{fig:fig3p5}, represented as $V$:
\begin{flalign*}
    &{A}_{1}^{i} = {V}_{i-1,i+1}, \\
\end{flalign*} 
After random permutation, the same bit holds a completely different covariance between attributes $j$ and $k$, determined by the permutation matrix:
\begin{flalign*}
    &{A}_{1}^{i} = {V}_{j,k}, \\
		&R_{1}(i-1)=j \text{ and } R_{1}(i+1)=k,\\
\end{flalign*}
where $R_1$ vector holds the random permutation of indices.\\
Therefore, a random permutation enables computation of a completely different set of attribute correlations. Effectively, random permutation and state evolution work in a similar fashion to include the computation of previously unseen attribute correlations. The equivalence of these two dimensions in CA framework is experimentally verified above (eg. Table \ref{table:5BitElem}). However, the effect of $R$ and $I$ is not completely symmetric for CA computation, and it is analyzed in the following sections. 

Cellular automaton feature space holds distributed higher order statistics of attributes, denoted as $A_k$. What happens if we use covariance features, $C_k$? Remember that these two can be transformed to each other, however do they differ in generalization performance? We repeated 5 bit memory task experiment (T=200, I=8) using $C_k$ features instead of $A_k$, and Table \ref{table:5BitElemCA_LocalRep} shows the comparison. Distributed representation stored as cellular automaton states show clear performance advantage over the local covariance representation, which emphasizes the importance of cellular automata based feature expansion. In the following sections (\ref{sec:kernellearning_section}) we compare polynomial SVM kernel with cellular automaton kernels (Table \ref{table:MNIST_KernelTwoFeatures}), and support this argument. Although the distributed higher order statistics argument we made through derivations were for linear cellular automata rules,  it can be extended for other interesting rules (eg. rule 110) by experimental studies, which is planned as a future study. As a side note, elementary cellular automata with memory (ECAM) framework provides a more powerful way of exploiting attribute statistics and it should be investigated both analytically and experimentally from reservoir computing perspective.

\begin{table}[h]                 
\centering                        
\begin{tabular}{l|l|l|l} 
$Rule 90, T=200, I=8$ & 16 & 32 & 64 \\  
\hline  
Distributed HOS & 73 & 9 & 0 \\       
Local Covariance & 94 & 62 & 11 \\
\end{tabular}                            
\caption{Percent failed trials for 5 Bit Task, $T_0 = 200$. Columns are number of random permutations. Distributed Higher Order Statistics (kept in CA states) and Local Covariance features are compared.}          
\label{table:5BitElemCA_LocalRep}        
\end{table}

\section{Metric Learning for Linear Cellular Automata Feature Space}
\label{sec:metriclearning_section}
Kernel based methods are widely used statistical machine learning tools \cite{vapnik2000nature} (see \cite{motai2014kernel} for a recent survey). The kernel trick enables implicit computation of instance distances (or dot products) in the expanded feature space, without ever computing the expanded feature. It significantly reduces the amount of computation in training and test stages, for many type of kernels, features and tasks. Kernelization is closely related with distance metric learning \cite{kulis2012metric,yang2006distance}, especially for nonlinear feature mappings. In the context of cellular automata feature space introduced in this paper, we would like to kernelize the system by devising a shortcut computation of cellular automata feature distances without cellular automata evolution computation. We show the kernelization for linear cellular automaton rule 90 due to its simplicity however this work can be extended for other analytically tractable rules. 

Linear cellular automaton state evolution can be represented as consecutive matrix multiplication \cite{martin1984algebraic, chaudhuri1997additive, ganguly2001theory}. Suppose we have CA initial conditions $A_0$ (vector of size $N$), $I$ number of state evolutions and $R$ number of random initial projections. Then, state evolution for a single random projection is given by, 
\begin{flalign*}
    &{A}_{1} = M_N * A_0 \text{  } mod \text{ } 2, \\
		&{A}_{2} = {M_N}^2 * A_0 \text{  } mod \text{ } 2, \\
		&{A}_{3} = {M_N}^3 * A_0 \text{  } mod \text{ } 2, \\
		&\text{. . .}\\
		&{A}_{I} = {M_N}^I * A_0 \text{  } mod \text{ } 2, \\
\end{flalign*}  
where $M_N$ is an $N\times N$ sparse binary matrix:
\[
M_N=
  \begin{bmatrix}
    0 & 1 & 0 & ... & 0 & 1\\        1 & 0 & 1 & 0 & ... &  0\\            0 & ... &  0 & 1 & 0 & 1\\        1 & 0 & ... &  0 & 1 & 0\\      
	\end{bmatrix}
\]
The feature space can be computed at once by concatenation of matrices:
\begin{flalign*}
	&M_I = [M_N \text{  } {M_N}^2 \text{  } {M_N}^3  \text{ ... }  {M_N}^I],\\
	&\mathbf{A} = M_I A_0 \text{  } mod \text{ } 2,
\end{flalign*} 
where $M_I$ is an $NI\times N$ matrix.

Since a random permutation of initial cellular automata configuration is another matrix multiplication, the feature space for a single random permutation is given by:
\begin{flalign*}
	&\mathbf{A^{R1}} = M_I  {\Pi}^{R_1} A_0 \text{  } mod \text{ } 2.
\end{flalign*} 
Then the cellular automaton feature space for all the random permutations (whole CA feature space, $\mathbf{A^R}$) can be computed using a single matrix multiplication:
\begin{flalign*}
	&\mathbf{M} = [M_I{\Pi }^{R_1} \text{  }  M_I{\Pi }^{R_2} \text{  } M_I{\Pi }^{R_3}   \text{ ... }  M_I{\Pi }^{R_R}],\\
	&\mathbf{A^R} = \mathbf{M} A_0 \text{  } mod \text{ } 2,
\end{flalign*}
where $\mathbf{M}$ is an $NIR\times N$ matrix, and $\mathbf{A^R}$ is a $NIR$ sized vector. 

Euclidian distance (equivalent to Hamming for binary data) between two data instances, $A_0$ and $B_0$, can be analyzed as follows:
\begin{flalign*}
	&\mathbf{d}(\mathbf{A^R},\mathbf{B^R}) =\|{\mathbf{M} A_0 \text{  } mod \text{ } 2 - \mathbf{M} B_0 \text{  } mod \text{ } 2}\|,\\
	& = \|{\mathbf{M} (A_0 - B_0) \text{  } mod \text{ } 2}\| \text{  } \tag*{because $-1^2=1^2$ , }\\
	& = (A_0-B_0)^T \text{  } \mathbf{M^T} mod \text{ } 2 \text{ } \times \text{ } \mathbf{M} (A_0-B_0) \text{  } mod \text{ } 2 \tag*{Euclidian distance property, }\\
	& = \left((A_0-B_0)^T  mod \text{ } 2 \text{ } \times \text{ } \mathbf{M^T} mod \text{ } 2\right) \text{  } mod \text{ } 2 \text{ }\\
	&\times \text{ } \left( \mathbf{M} \text{ } mod \text{ } 2  \text{ } \times \text{ } (A_0-B_0) \text{ } mod \text{ } 2 \right) \text{  } mod \text{ } 2 \tag*{Property of modulus, }
\end{flalign*}
If we take the modulus 2 of both sides of the equation,
\begin{flalign*}
	&\mathbf{d}(\mathbf{A^R},\mathbf{B^R}) \text{  } mod \text{ } 2 \text{ } = (F \text{  } mod \text{ } 2 \text{ } \times G \text{  } mod \text{ } 2 \text{ }) \text{  } mod \text{ } 2 \text{ }, 
	& 
\end{flalign*}
where $F$ and $G$ are new variables for the complex terms in multiplication given in the previous equation. Then we can use the property of modulus again to reach:
\begin{flalign*}
	&\mathbf{d}(\mathbf{A^R},\mathbf{B^R}) \text{  } mod \text{ } 2 \text{ } = (F \times G) \text{  } mod \text{ } 2 \text{ }. 
\end{flalign*}

Unwrapping $F$ and $G$, we get
\begin{flalign*}
	&\mathbf{d}(\mathbf{A^R},\mathbf{B^R}) \text{  } mod \text{ } 2 \text{ } = \\
	&\left((A_0-B_0)^T  mod \text{ } 2 \text{ } \times \mathbf{M^T} mod \text{ } 2 \times \mathbf{M} \text{ } mod \text{ } 2  \text{ } \times (A_0-B_0) \text{ } mod \text{ } 2\right) \text{ } mod \text{ } 2  \text{ }. 
\end{flalign*}
Let us define two new variables for simplification:
\begin{flalign*}
	&d_0 = (A_0-B_0) \text{ } mod \text{ } 2 \text{ }\\
	&\mathbf{M_K}=\mathbf{M^T} mod \text{ } 2 \times \mathbf{M} \text{ } mod \text{ } 2,
\end{flalign*}
then we attain,
\begin{flalign*}
	&\mathbf{d}(\mathbf{A^R},\mathbf{B^R}) \text{  } mod \text{ } 2 \text{ } = {d_0}^T \text{ } \mathbf{M_K} {d_0} \text{ } mod \text{ } 2  \text{ }.
\end{flalign*}

Assuming that we can estimate the manifold via a function ($g$) defined on modulus 2, we get a definition of distance,
\begin{flalign*}
	&\mathbf{d}(\mathbf{A^R},\mathbf{B^R}) = {d_0}^T \text{ } \mathbf{M_K} {d_0} \text{ } + \text{ } 2 \times \mathbf{g}(d_0,\mathbf{M_K})\\
	&\mathbf{d}(\mathbf{A^R},\mathbf{B^R}) = \mathbf{f}(d_0,\mathbf{M_K}).
\end{flalign*}

\begin{figure*}
\begin{center}
\includegraphics[width=1\textwidth]{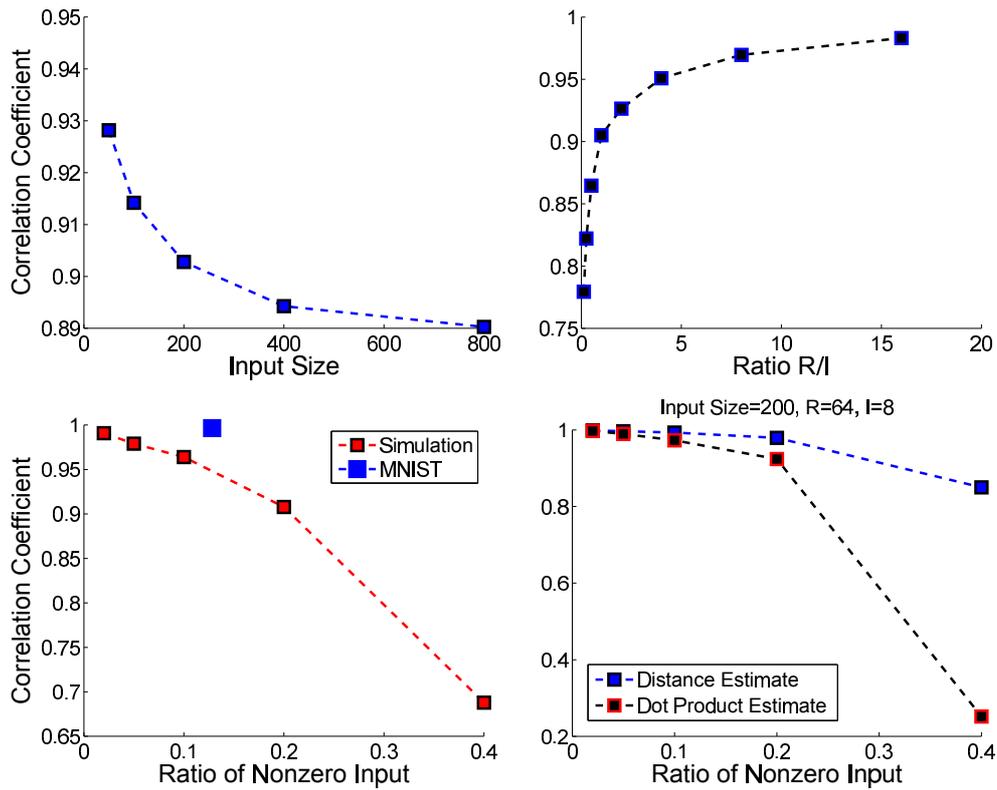} 
\end{center}
   \caption{Results on metric learning simulations. The distance between data instances are estimated via linear regression using a set of relevant features. Average correlation coefficient is presented with respect to a set of factors: input size $N$, ratio of number of permutations and CA time evolutions $R/I$, ratio of non-zero elements in the input array $Nz$. The results are average of a set of factor combinations. Lower right plot shows the result for a specific configuration for both distance estimation and dot product estimation (see section \ref{sec:kernellearning_section}). }
\label{fig:fig4}
\end{figure*}

The function $f$ estimates the CA feature distances (i.e. $\mathbf{d}(\mathbf{A^R},\mathbf{B^R})$) by utilizing the distance between the initial CA states of two instances ($d_0$), the matrix $\mathbf{M_K}$ of size $N\times N$, which is specified by the random initial projections and the number of CA iterations. A few observations:\\
1. ${d_0}^T{d_0}$ is the Euclidian (Hamming) distance between the data instances\\
2. ${d_0}^T \text{ } \mathbf{M_K} {d_0}$ is analogous to Mahalanobis distance with covariance matrix $\mathbf{M_K}$.\\
Can we estimate function $f$ for \textbf{a given dataset}? The experiments reported below (Figure \ref{fig:fig4}) indicate that it is possible to very accurately estimate the distance using linear regression over features of $d_0$ and $\mathbf{M_K}$. But should we use the entries of $d_0$ and $\mathbf{M_K}$ as features, or can we devise more meaningful features? We conjecture that, the distance in CA feature space will depend on both the initial Euclidian and Mahalanobis distances of the data. 


Another alternative is usage of eigenvalue matrix $\mathbf{S_K}$, derived by singular value decomposition of $\mathbf{M_K}$, for weighted Euclidian distance computation:
 ${d_0}^T \text{ } \mathbf{S_K} \text{ } {d_0}$. 
It can be computed in $\mathcal{O}(N)$ time due to the diagonal nature of $\mathbf{S_K}$. Mahalanobis distance is not only computationally more demanding but also overfitting the densities, producing unnecessarily large and elongated hyperellipsoidal components \cite{mao1996self}, hence it is a good idea to regularize it by the eigenvalue matrix \cite{archambeau2004flexible}. Usage of the three distance features for linear regression of CA feature distance, will effectively perform that regularization. Let us summarize the features for regression:
\begin{flalign*}
	&\mathbf{e_1} = {d_0}^T \text{ }  {d_0} \\
	&\mathbf{e_2} = {d_0}^T \text{ } \mathbf{M_K} {d_0} \\
	&\mathbf{e_3} = {d_0}^T \text{ } \mathbf{S_K} {d_0}
\end{flalign*} 

We ran a Monte Carlo simulation \footnote{20 simulations per parameter combination.} to measure the quality of CA feature distance estimation according to a set of factors: size of data vector ($N$), ratio of nonzero elements ($Nz$), number of random permutations ($R$), number of CA evolutions ($I$). The distance features [$\mathbf{e_1}; \mathbf{e_2}; \mathbf{e_3}$] are used for linear regression \footnote{Linear regression was performed with pseudo inverse on the data matrix.} of the CA feature distances of two data instances. A hundred random binary vectors are created, and their pairwise CA feature distances as well as the distance features are computed. $80\%$ of the CA feature distances ($3960$ pairs) are used for training the regressor, the rest of the pairs ($990$) are used for the test. The results are given in Figure \ref{fig:fig4}. The mean correlation between the real distance and the estimate is:\\
1. moderately decreasing with vector size, leveling around $400$ size inputs, \\
2. sharply increasing with $R/I$ ratio, leveling around $R=5I$,\\
3. linearly decreasing with ratio of nonzero elements.

The ratio $R/I$ and ratio of nonzero elements determine the estimation quality, and we propose that it is due to the increased Kolmogorov complexity (uncompressibility) of the state evolutions in time \cite{zenil2013asymptotic}.

The experiments indicate that, given an input set, it is possible to very accurately estimate CA feature distance metric with proper choice of parameters and preprocessing (subsampling explained in the next section). The data used in the simulations did not have any structure, yet it is possible to learn distances, which suggests that the major inherent factor in the learned metric is the ratio of nonzero elements ($Nz$) and vector size ($N$). We also tested the metric on MNIST handwritten digit recognition dataset (Figure \ref{fig:fig4}, lower left). The first $200$ digits were used to evaluate the metric learning performance, in which there is a large variance (mean $0.13$ and std $0.04$) on the ratio of nonzero elements \footnote{$R=64$ and $I=8$}. The correlation coefficient is very high despite the variance in $Nz$, which suggests that the learned metric is robust also for structured data.

\section{Linear Complexity Kernel for Support Vector Machine Learning}
\label{sec:kernellearning_section}
We want to derive the kernel function for dot product/projection of CA features, that is intended to be used in support vector machines. The form of the proposed kernel derived through dot product in CA feature space enables the exchange of summation trick used in support vector machine linear kernels, that reduces computation time from $\mathcal{O}(DN)$ ($D$ is the number of support vectors) to $\mathcal{O}(N)$. Given the fact that interesting problems require thousands of support vectors \cite{cao2008approximate}, this corresponds to a \textbf{threefold speedup}. 

The projection of one CA feature over another is:
\begin{flalign*}
	&(\mathbf{A^R})^T .\text{ } \mathbf{B^R} = {A_0}^T \text{ } \mathbf{M}^T \text{  } mod \text{ } 2 \text{ } . \text{ } \mathbf{M} \text{ } B_0 \text{  } mod \text{ } 2 \text{ }. 
\end{flalign*}
Using the property of modulus and the steps similar as before  we get,
\begin{flalign*}
	&(\mathbf{A^R})^T .\text{ }\mathbf{B^R} \text{  } mod \text{ } 2 \text{ } = {A_0}^T \text{ } \mathbf{M_K}  \text{ } B_0  \text{  } mod \text{ } 2 \text{ }. 
\end{flalign*}

It is possible to estimate the projection via learning a function discussed in the previous section:
\begin{flalign*}
	&(\mathbf{A^R})^T .\text{ }\mathbf{B^R} = {A_0}^T \text{ } \mathbf{M_K}  \text{ } B_0  + \text{ } 2 \times \mathbf{g}({A_0}^T . \text{ } B_0,\mathbf{M_K})\\
	&(\mathbf{A^R})^T .\text{ }\mathbf{B^R} = \mathbf{f}({A_0}^T . \text{ } B_0,\mathbf{M_K}).
\end{flalign*}
The initial projection ${A_0}^T \text{ } B_0$, weighted initial projection ${A_0}^T \text{ } \mathbf{S_K}  \text{ } B_0$ and 'Mahalanobis' projection ${A_0}^T \text{ } \mathbf{M_K}  \text{ } B_0$ can be used as the features (similar with $e1$, $e2$, $e3$ above) for learning the estimation function of the CA feature projection. Weighted initial projection is nothing more than a vector product of the form, $({A_0}^T \text{ } .* \text{ }\mathbf{s})   \text{ } B_0$, in which $.*$ is elementwise product and $\mathbf{s}$ is the diagonal vector of $\mathbf{S_K}$. The quadratic term requires $\mathcal{O}(N^2)$ time due to matrix vector multiplication, but it will be performed offline only once during training, which is explained below. 

Metric learning with linear regression gives three coefficients, and the projection estimate is given by,
\begin{flalign*}
	&(\mathbf{A^R})^T .\text{ }\mathbf{B^R} = k_1 .\text{ } {A_0}^T . \text{ } B_0 + k_2 .\text{ } ({A_0}^T \text{ } .* \text{ }\mathbf{s})   \text{ } B_0 + k_3 .\text{ } {A_0}^T \text{ } \mathbf{M_K}   \text{ } B_0.
\end{flalign*}
The performance of the three metrics on estimation of CA feature similarity is given in Figure \ref{fig:fig4}, lower right plot. It is observed that dot product estimate is as good as distance estimate, but breaks down for dense vectors (large ratio of nonzero elements). However if the data is dense, the detrimental effect on dot product estimation can always be avoided by subsampling during the random permutation initialization stage of the algorithm. \footnote{This will increase the computational demands, i.e. larger $R$ and $I$ will be neccessary. However, this is not an issue for the kernel method, because kernel matrix $\mathbf{M_K}$ will be computed once, offline. }

Then in support vector machine (SVM) framework, we can use the newly defined kernel. The prediction function of an SVM is given by,
\begin{flalign*}
	& w^T . \text{ } {X^R} = \displaystyle\sum_{i=1}^{D} \alpha_i \text{ } y^{(i)} k(X,{\mathbf{Y}}^{(i)}),
\end{flalign*}
where $w$ is the weight, ${X^R}$ is the input representation in CA feature space. This costly multiplication due to high dimensionality is replaced using the summation in the right hand side. $D$ is the number of support vectors and $\mathbf{Y}$ is the matrix that holds them, $y^{(i)}$ is the output value (class label or regressed variable) of the support vector, $\alpha_i$ is the weight, $k$ is the kernel function, and $X$ is the input raw data. Using the kernel obtained via linear regression over the defined features, we get,
\begin{flalign*}
	& w^T . \text{ } {X^R} = \displaystyle\sum_{i=1}^{D} \text{ } \alpha_i \text{ } y^{(i)} \text{ } \displaystyle\sum_{j=1}^{N} ( \text{ } k_1 \text{ }.\text{ } {X_j} {\mathbf{Y}}^{(i)}_j \text{ } + \text{ } k_2  \text{ } . \text{ } {X_j} \text{ } {\mathbf{U}}^{(i)}_j \text{ } + \text{ } k_3 \text{ } {X_j} {\mathbf{Z}}^{(i)}_j \text{ })
\end{flalign*}
where ${\mathbf{U}}^{(i)} = \mathbf{s} \text{ } .* \text{ } {\mathbf{Y}}^{(i)}$ and ${\mathbf{Z}}^{(i)} = \mathbf{M_K} \text{ }  {\mathbf{Y}}^{(i)}$  .

Exchanging the place of summation terms and rearranging,  
\begin{flalign*}
	& w^T . \text{ } {X^R} = \displaystyle\sum_{j=1}^{N} \text{ } {X_j}  \text{ } \displaystyle\sum_{i=1}^{D} \alpha_i \text{ } y^{(i)} \text{ }   (\text{ } k_1 \text{ } {\mathbf{Y}}^{(i)}_j \text{ } + \text{ }  k_2  \text{ } . \text{ }  {\mathbf{U}}^{(i)}_j \text{ } + k_3 \text{ } {\mathbf{Z}}^{(i)}_j  \text{ }).
\end{flalign*}
The second summation (denoted as $Q_j$) can be computed offline and saved, to be used during both training and test:
\begin{flalign*}
	& w^T . \text{ } {X^R} = \displaystyle\sum_{j=1}^{N} \text{ } {X_j} \text{ } . \text{ } Q_j.\\
\end{flalign*}

There are three remarks about the kernelization of CA feature for linear rules: 

1. Two new types of support vectors are defined ${\mathbf{U}}^{(i)}$ and  ${\mathbf{Z}}^{(i)}$, and kernel is a weighted combination of the three support vectors, which resembles \cite{jebara2004multi}. 

2. Standard echo state networks are kernelized in \cite{hermans2012recurrent,shi2007support}. These studies are enlightening because they bridge the gap between kernel based machine learning and recurrent neural networks. Yet the computational complexity of the proposed recursive kernelizations in these studies is at least as large as nonlinear kernel methods. 

3. There are studies for imitating nonlinear kernels via polynomial approximations \cite{cao2008approximate,claesen2014fast}. They transform the computational complexity from $\mathcal{O}(DN)$ to $\mathcal{O}(N^2)$, which gives speedups only for low dimensional data. 

However, different from the previous studies, \textbf{the computational complexity of the cellular automata kernel is identical to standard linear kernel, $\mathcal{O}(N)$.} Yet, linear cellular automata rules are known to be very powerful computational tools, comparable to Turing complete rules (see Tables \ref{table:5BitElem} and \ref{table:MNIST_KernelTwoFeatures}, \cite{baetens2010phenomenological,martinez2013note,alonso2006elementary,zenil2013asymptotic}). 

\begin{table}[h]                 
\centering                        
\begin{tabular}{l|l|l|l|l} 
 Dataset & Linear Raw  & CA Kernel & RBF Kernel & Polynomial Kernel \\  
\hline  
MNIST & 85 & 86  & 86.5 & 82.5   \\
\end{tabular}                            
\caption{Performance comparison of classifiers on subset of MNIST dataset. See text for details.}          
\label{table:MNIST_KernelTwoFeatures}        
\end{table} 

How does the CA kernel perform on structured data? We compared \footnote{The experiments were not performed using the whole datasets or large $(R,I)$ combinations due to memory issues. Memory efficient implementation is necessary because off-the-shelf libraries are not fit for our framework.  A comparison with state-of-the-art is planned for the near future. } it with other SVM kernels \cite{CC01a} on MNIST dataset \cite{lecun1998gradient}. We used the first $200$ data instances (binarized with intensity threshold $128$) to learn the kernel function (two features). We used the kernel function to get the kernel matrix (pairwise projections between all data instances), then trained a classifier using the estimated kernel matrix. We tested the classifier on the next $200$ data points again using the estimated kernel matrix. The results are given in Table \ref{table:MNIST_KernelTwoFeatures} where we compare linear , CA ($R=128$ and $I=8$), RBF and polynomial (order $2$) kernels. Coarse-to-fine grid search is done on parameter space for finding the best performance for all classifiers. It is observed that CA kernel approaches RBF kernel, even though the computational complexity is linear. Interestingly, polynomial kernel severely overfits for orders higher than $2$, yet the CA kernel estimates $8^{th}$ order statistics (because $I=8$) but does not overfit which shows a clear advantage of distributed representation (see also Table \ref{table:5BitElemCA_LocalRep}). Having said that, the true power of CA computation is tapped only when hybrid/nonlinear rule based, multi-layered CA feature spaces are built that utilize unsupervised pre-training (see Discussion section for potential improvements).

\begin{figure*}
\begin{center}
\includegraphics[width=1\textwidth]{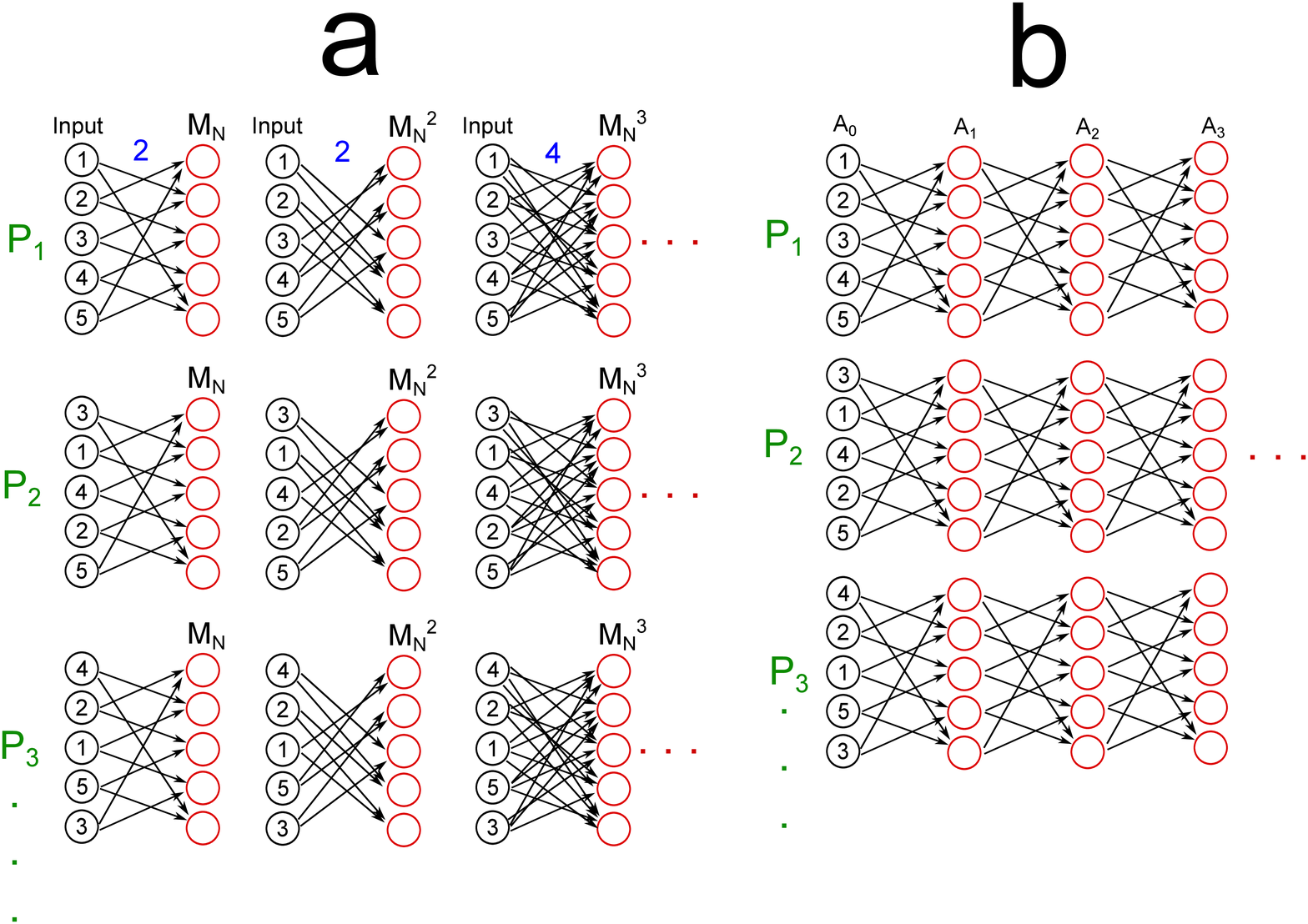} 
\end{center}
   \caption{The network formed by the CA feature space of rule 90 CA. \textbf{a.} The network can be viewed as a collection of connected components, which is composed of different input permutations (P, in rows) and time unrolled CA evolutions (in columns). The average number of connections per unit increases with time, as shown in blue in the uppermost row. \textbf{b.} The network can be viewed as a time unrolled feedforward network, however the connections are not all-to-all between layers due to partitioning of different permutations, given as separate rows. }
\label{fig:fig5}
\end{figure*}

\section{Analysis of the Feeforward Cellular Automata Network}
\label{sec:NetworkArchitecture}
The cellular automaton feature space can be perceived as a network, an illustration for rule 90 is given in Figure \ref{fig:fig5} ($N=5$). In one view (\textbf{a} in the figure), the network consists of $R\times I$ number of connected components shown in a grid. In this representation, the time dimension in CA evolution is \textbf{unrolled} and given in the columns, where the adjacency matrix for time $k$ is determined by $M_N^k \text{  } mod \text{  } 2$ \footnote{Modulus operation simplifies $M_N$ because even entries are ineffective and they are zeroed out after the modulus.}. Each permutation is a copy of the time unrolled network, that receives from a different permutation of inputs, given as rows. For simplicity of illustration, the input projection (the indices 1,2,3,4,5 are shown for input vector) is demonstrated separately for each connected component. The average number of connections increases with the time evolution (see blue numbers below the connected components in the first row), because $M_N^k \text{  } mod \text{  } 2$ density increases with $k$. This observation is compatible with the Kolmogorov complexity experiments in \cite{zenil2013asymptotic}, that state the increase in uncompressibility of CA states in time. One other perspective on the uncompressibility is related with the pattern replication property of rule 90 \cite{wolfram1983statistical}. As time evolves, it is less likely to replicate a portion of the initial pattern, due to increased number of interactions (i.e. more connections with increasing $I$ in the network, in Figure \ref{fig:fig5}), hence reduce the probability of effective data compression. 

There are $N \times I \times R$ neurons in this feedforward network. The neurons can be arranged to obtain the a recurrent neural network equation in classical form. Suppose $A_0^R$ is a $N\times R$ size matrix which holds the initial conditions (i.e. raw data) of the CA for $R$ different permutations. Then,
\begin{flalign*}
	& A_1^R = M_N A_0^R \text{  } mod \text{  } 2 \\
	& A_2^R = M_N A_1^R \text{  } mod \text{  } 2  \\
	& ...\\
	& A_k^R = M_N A_{k-1}^R \text{  } mod \text{  } 2 ,
\end{flalign*} 
where $M_N$ is a the characteristic matrix defining state evolution of the linear CA defined in section \ref{sec:metriclearning_section}. The recursive computation definition of recurrent neural network is in autonomous mode (i.e. zero input), and this is due to flattening of the input. Using this recurrent formulation, CA expansion can also be viewed as a time unrolled continuous (i.e. not connected components) network shown in  Figure \ref{fig:fig5} b. 

\textbf{How about nonlinear CA rules?} We can write a generalized the CA network equation as follows:
\begin{flalign*}
	& A_k^R = \mathbf{f} (A_{k-1}^R) ,
\end{flalign*} 
where function $\mathbf{f}$ is over the neighbors of the cells and it approximates the rule of the CA. If the rule is not 'linearly separable' then we have to use multiple layers. Therefore, each time step in Figure \ref{fig:fig5} b might require a number of hidden layers. The overall network is a repetition (at each time step) of a multi-layer perceptron, with heaviside step nonlinearity at the last layer. To the best of our knowledge, this is a novel perspective, and 'neuralization' of the CA rules seems like an interesting research direction.

\section{Cellular Automata Reservoir}
\label{sec:CA_Reservoir_Recurrent}
In feedforward CA feature expansion experiments we presented the sequence as a whole by flattening: whole input sequence is vectorized, CA feature is computed, then whole output feature is estimated using this large holistic sequence representation. Using a larger data portion for context is common in sequence learning and it is called sliding window method, and estimating the whole output sequence is generally utilized in hidden Markov model approaches \cite{dietterich2002machine}, but these two were not applied together before to the best of our knowledge. We adopted this approach for 5/20 bit memorization tasks, however it is possible to estimate each time step in the output sequence one at a time, as in echo state networks. Suppose that we would like to estimate $O_t$, the output at time step $t$. Using Markovian assumption we say that only the input up to $M$ steps back is relevant, then we use the chunk of the input sequence $(I_{t-M}; I_{t-M-1}; ... I_{t-1})$ as an input to the cellular automata reservoir. This assumption is widely used in classical recurrent neural network studies, in which the hidden layer activities are reset at every $M$ time steps (eg. see Language Modeling in \cite{pascanu2012difficulty}). We use this chunk of input as initial conditions of the cellular automata, compute CA evolution and use CA reservoir feature vector for regression/classification. Why didn't we use this 'classical' approach in 5 bit and 20 bit tasks in section \ref{sec:results_section}? We have to discard the first $M$ output time steps because we need $M$ input time steps to estimate the ${M+1}^{st}$ time step of the output sequence. However, if we had discarded the first $M$ time steps, it wouldn't be a fair comparison with results on echo state networks. Please note that, estimating the whole sequence using the whole input sequence is a much harder task, and we are making our case more difficult in order to keep the comparison with literature. 

Let us emphasize that using the Markovian approach, CA reservoir is able to perform any task that echo state network (or recurrent neural network in general) is able to do. Yet, we have to choose the right amount of history needed for estimation (i.e. parameter $M$). The size of the representation increases with $M$ in our algorithm. The advantage of echo state networks over our approach is its capability to keep a fixed size representation (activity of neurons in the reservoir) of timeseries history of arbitrary size. However, as the size of the timeseries history needed to perform the task increases, the size of the reservoir network should increase proportionally, suggested by the experiments in \cite{jaeger2012long}. Then for feedforward cellular automata expansion there is a selection of history $M$ that constitutes the feature space, and for echo state networks the number of neurons that accommodate timeseries history of $M$ should be decided. 

Feedforward cellular automata is shown to be capable of long-short-term-memory. Nevertheless we would like formulate a classical recurrent architecture using cellular automata feature expansion, because it is a much intuitive representation for sequential tasks. 

Suppose that cellular automaton is initialized with the first time step input of the sequence, $X_0$. The vector ${X_0}^{P_1}$ is a single random permutation of the input. In contrast to feedforward formulation, we concatenate multiple random permutations to form a $N \times R$ vector in the beginning of CA evolution:
\begin{flalign*}
		&{X_0}^{P} = [{X_0}^{P_1}; {X_0}^{P_2}; {X_0}^{P_3}; ... {X_0}^{P_R}]
\end{flalign*}

Cellular automaton is initialized with ${X_0}^{P}$ and evolution is computed using a prespecified rule, $Z$ (Figure \ref{fig:fig3}):
\begin{flalign*}
    &{A_1} = Z({X_0}^{P}), \\
		&{A_2} = Z({A_1}), \\
		&...\\
		&{A_I} = Z({A_{I-1}}).
\end{flalign*}  
$A_1$ up to $A_I$ are state vectors of the cellular automaton. We concatenate the evolution of cellular automata to obtain a single state vector of the reservoir (size $NIR$), to be used for estimation at time $1$: 
\begin{flalign*}
		&{A}^{(1)} = [{A_1}; {A_2}; ... {A_I}].
\end{flalign*}
We have to insert the next time step of the input, $X_1$ into the reservoir state vector. There are many options, but here we adopt normalized addition of state vector and input vector \cite{kanerva2009hyperdimensional}, in which, entries with value $2$ (i.e. $1+1$) become $1$, with value $0$ stay $0$ (i.e. $0+0$), and with value $1$ (i.e. $0+1$) are decided randomly. We modify the state vector of the cellular automaton at time $I$:
\begin{flalign*}
	  &{A_I} = [{A_I} + {X_1}^{P}],
\end{flalign*}  
in which square brackets represent normalized summation. The cellular automaton is evolved for $I$ steps to attain ${A}^{(2)}$,
\begin{flalign*}
		&{A}^{(2)} = [{A_{I+1}}; {A_{I+2}}; ... {A_{2I}}],
\end{flalign*}
which is used for estimation at time step $2$. This procedure is continued until the end of the sequence, when $X_T$ is inserted into the reservoir.   

We analyzed the recurrent evolution for additive rule 90 \footnote{For simplification $I$ is assumed to be a power of 2.}, as we did in section \ref{sec:distributedRep}. Interestingly, the recurrence at multiples of $I^{th}$ time step is the same with heteroassociative storage of sequences proposed in \cite{kanerva2009hyperdimensional,gallant2013representing,plate2003holographic}:
\begin{flalign*}
	  &{A_I} = {\Pi}_{I} {X_0}^{P} \oplus \text{  } {\Pi}_{I} {X_0}^{P}, \\
		&{A_{2I}} = {\Pi}_{I} ({X_1}^{P} \oplus \text{  } {\Pi}_{I} {X_0}^{P}) \text{  } \oplus \text{  }  {\Pi}_{-I} ({X_1}^{P} \oplus \text{  } {\Pi}_{-I} {X_0}^{P}) \\
		&{A_{3I}} = {\Pi}_{I} ({X_2}^{P} \oplus \text{  } {\Pi}_{I} ({X_1}^{P} \oplus \text{  } {\Pi}_{I} {X_0}^{P})) \text{  } \oplus \text{  }      {\Pi}_{-I} ({X_2}^{P} \oplus \text{  } {\Pi}_{-I} ({X_1}^{P} \oplus \text{  } {\Pi}_{-I} {X_0}^{P}))\\
		& ...
\end{flalign*}  

However, different from \cite{kanerva2009hyperdimensional,gallant2013representing,plate2003holographic}, we are using the whole set of cellular automaton state vectors (i.e. $I$ time steps of evolution) as the reservoir, and it can used for estimating the sequence output via any machine learning algorithm (SVM, logistic regression, multilayer perceptron etc.) other than autoassociative retrieval. Nevertheless, additive cellular automata is very similar to a linear recurrent architecture proposed by Kanerva, Gallant and previously by Plate. Yet, there is a large arsenal for cellular automata rules, most of which are nonlinear and some of them Turing complete. Then our approach is a generalization of previously proposed recurrent architectures, in terms of atomic operation (i.e. linear vs nonlinear), readout (nearest neighbor retrieval vs a wide set of machine learning algorithms) and most importantly the representation (random vector/matrix vs cellular automata states).   

\begin{figure*}
\begin{center}
\includegraphics[width=0.5\textwidth]{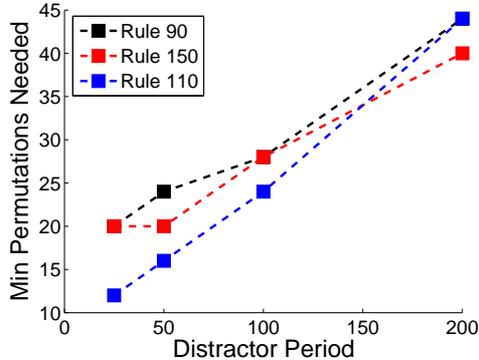} 
\end{center}
   \caption{Distractor period vs the minimum number of permutations needed to obtain zero error in 5 bit memorization task is given for three different elementary cellular automata rules. }
\label{fig:fig_dyn}
\end{figure*}

We tested the performance of the cellular automata reservoir on 5 bit task. We used $I=32$, and varied the number of permutations $R$, to obtain the minimum $R$ value (y axis in Figure \ref{fig:fig_dyn}) to get zero error in 5 bit tasks of various distractor periods (x axis in Figure \ref{fig:fig_dyn}). In Figure \ref{fig:fig_dyn}, we are showing the results for three cellular automata rules. It is observed that reservoir size (i.e. $R \times I$) demand has increased compared to feedforward cellular automata feature expansion (Table \ref{table:5BitElem}). Yet memory requirements in training stage is much smaller, due to the abandonment of flattening. Most importantly, recurrent architecture enables generation of unbounded-length signal output, and acceptance of unbounded-length timeseries input. 

\textbf{Can we generalize our analyses on feedforward CA feature expansion for CA reservoir?} Cellular automaton evolution and feature computation after insertion of a sequence time step (${A}^{(1)}$ up to ${A}^{(T)}$) is equivalent to feedforward computation given in section \ref{sec:method_section}. Then we can say that:

1. The cellular automaton states hold a distributed representation of attribute statistics, as well as correlation between sequence time steps (section \ref{sec:distributedRep}).

2. The reservoir feature vector can be kernelized for efficient implementation in support vector machine framework (section \ref{sec:kernellearning_section}). 

3. The CA reservoir can be viewed as a time unrolled recurrent network with xor connections instead of algebraic synapses (section \label{ref:NetworkArchitecture}).

\begin{figure*}
\begin{center}
\includegraphics[width=1\textwidth]{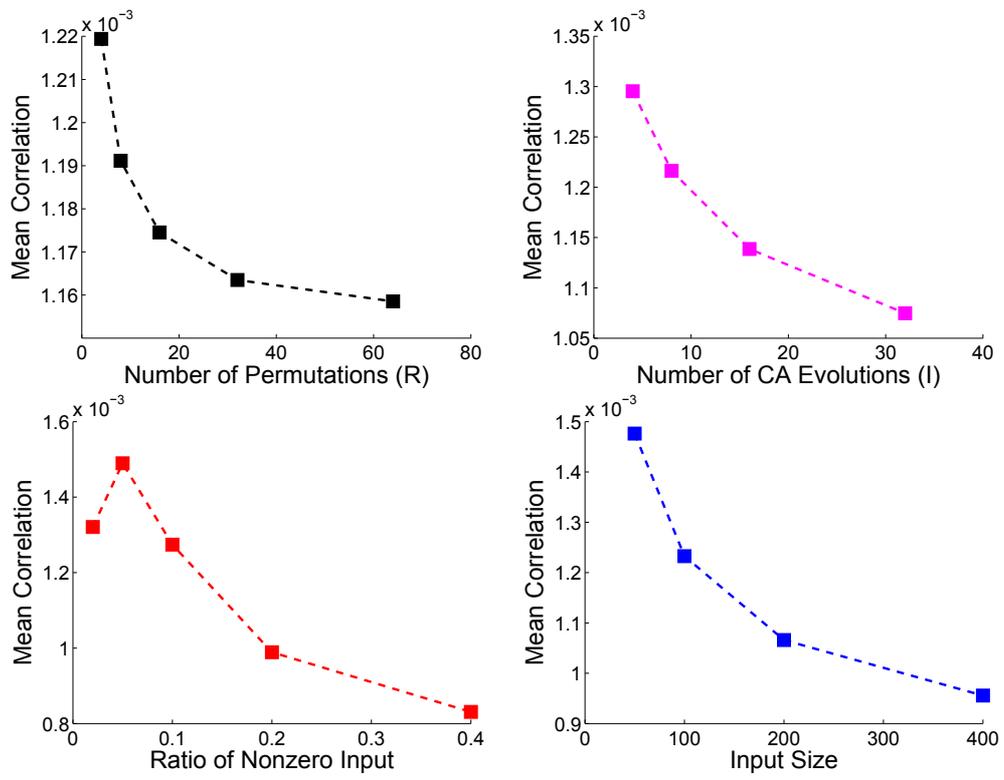} 
\end{center}
   \caption{The simulation results that measures the mean pairwise correlation of the CA features for different inputs. A smaller number is an indicator of more computational power. }
\label{fig:fig6}
\end{figure*}

\section{Computational Power}
The computational power of a system has many aspects, and it is a matter under debate for both cellular automata \cite{martinez2013note} and reservoir computing \cite{lukovsevivcius2009reservoir}. The experiments on 5 bit and 20 bit noiseless memory tasks (Table \ref{table:5BitGoL200and1000}) suggest that the capability increases with both number of permutations $R$ and number of CA evolutions $I$. We also wanted to test other effects such as size of data vector ($N$), ratio of nonzero elements ($Nz$) as in section \ref{sec:metriclearning_section}. In a simulation for rule 90, we computed mean pairwise correlation between CA features of randomly created vectors. The smaller this is, the larger the expected computational power \cite{lukovsevivcius2009reservoir}. The results are shown in Figure \ref{fig:fig6}, which indicate that computational power increases with vector size and ratio of nonzero elements. We will extend these simulations for all elementary rules (including ECAM) and other metrics in a future work.

\section{Computational Complexity}
There are two major savings of cellular automata framework compared to classical echo state networks:

1. Cellular automaton evolution is governed by bitwise operations instead of floating point multiplications. 

2. Since the CA feature vector is binary, matrix multiplication needed in the linear classification/regression can be replaced by summation. 

Multiplication in echo state network is replaced with bitwise logic (eg. XOR for rule 90) and multiplication in classification/regression is replaced with summation. Overall, multiplication is completely avoided, both in CA feature computation and in classifier/regression stages, which makes the CA framework especially suitable for FPGA hardware implementations. 

\begin{table}[h]                 
\centering                        
\begin{tabular}{l|l|l|l} 
Task & ESN (Floating Point) & CA (Bitwise) & Speedup  \\  
\hline     
5 bit $T_0=200$ & 1.03 M & 0.43 M & 2.4X  \\
5 bit $T_0=1000$ & 13.1 M & 8.3 M  & 1.6X \\  
20 bit $T_0=200$ & 17.3 M & 9.5 M  & 1.8X \\   
\end{tabular}                     
\caption{The comparison of the number of operations for the echo state networks (ESN) and the Cellular Automata (CA) framework. Operation is floating point for ESN, but bitwise for CA.}
\label{table:ComparisonOperation}        
\end{table}

The number of operations needed for the CA feature computation of 5 bit and 20 bit tasks is given in Table \ref{table:ComparisonOperation}, both for Echo State Network (ESN) in \cite{jaeger2012long} and for cellular automata (only feedforward architecture shown in the table). \footnote{For ESN, it is assumed that the number of floating point operations is equal to 2*NNZ.} There is a speedup in the number of operations in the order of 1.5-3X. However, considering the difference of complexity between floating point and bitwise operations, there is \textbf{almost two orders} of magnitude speedup/energy savings. \footnote{CPU architectures are optimized for arithmetic operations: bitwise logic takes 1 cycle and 32 bit floating point multiplication takes only 4 cycles, on 4th generation Haswell Intel\textsuperscript{TM} core. Therefore the speedup/energy savings due to the bitwise operations will be much more visible on hardware design, i.e. FPGA.} When looked into the CA reservoir, it is observed that about 3-4 times more computation is needed compared to feedforward feature expansion, yet it is still at least an order of magnitude faster than neuron based approaches.  

When support vector machines are used for linear cellular automaton rule kernels, the computational complexity is $\mathcal{O}(N)$, in which $N$ is the size of the initial data vector. However computational complexity is $\mathcal{O}(NIR)$ for general rules, where $R$ is the number of permutations and $I$ is the number of CA evolutions \footnote{Because linear methods are applied on CA feature space of size $NIR$.}. Given the fact that interesting problems would require $R>30 \text{ and } I>30$ (eg. Table \ref{table:20BitGoL200and300}), then usage of linear CA rule kernels will provide at least \textbf{three orders of magnitude} speedups \footnote{For 20 bit task, $R \times I \approx 6000$.}. The same argument is valid when we compare CA kernel with nonlinear kernels (eg. RBF), where the complexity is $\mathcal{O}(DN)$, where $D$ is the number of support vectors in the order of thousands for interesting problems.

\section{Symbolic Processing and Non-Random Hyperdimensional Computing}
\subsection{Combining Connectionist and Symbolic Capabilities}
\label{sec:CombineConnectionistSymbolic}
In this section, we would like to demonstrate the symbolic computation capability of the cellular automata reservoir. The starting point is the idea that hidden layer neural activities can be used as fixed-length embeddings of a wide variety of data, such as vision or language. One of the main topics of current AI research is to learn these embeddings from the data. On the other extreme, even random embeddings can be used to do symbolic computation as in reduced representation or vector symbolic architectures. We believe that, it is not a good idea to learn the embedding from the data/task since it is expected to degrade the generalization capability for unseen data and a different task \cite{gallant2013representing}. Also, it might not be necessary for performance as long as you use a computationally powerful feature expansion, i.e. cellular automata. Yet, random embedding is not very useful if one needs to use classical statistical machine learning tools, such as support vector machines \footnote{The main inference algorithm for random embedding approaches is 1 nearest neighbor matching, also called clean-up or autoassociative memory. This shortcoming is due to the fact that randomness does not allow to build a unitary model (eg. a decision boundary, a regression weight matrix etc.), as in statistical machine learning approaches.}. This shortcoming was spotted by Gallant et al.: "It is possible to improve recognition ability for bundle vectors when the vectors added together are not random..." \cite{gallant2013representing}.  Cellular automata reservoir gives a balanced approach: skip embedding learning, but keep the potential for machine learning. 

Two relevant symbolic processing frameworks are Conceptors by Jaeger \cite{jaeger2014controlling} and Hyperdimensional Computing by Kanerva \cite{kanerva2009hyperdimensional}. Conceptors use neural representation harvested via a neural reservoir, whereas Hyperdimensional Computing utilizes binary vectors. Although both of the approaches can be pursued for cellular automata based reservoir computing, in this study we are exploring the expressive power of Hyperdimensional representation. 

Hyperdimensional computing uses \textbf{random} very large sized binary vectors to represent objects, concepts and predicates. Then appropriate binding and grouping operations are used to manipulate the vectors for hierarchical concept building, analogy making, learning from a single example etc, that are hallmarks of symbolic computation. The large size of the vector provides a vast space of random vectors, two of which are always nearly orthogonal. Yet, the code is robust against a distortion in the vector due to noise or imperfection in storage, because after distortion it will still stay closer to the original vector, than the other random vectors. 

The grouping operation is normalized vector summation and it enables forming sets of objects/concepts. The resultant vector is similar to all the elements of the vector. The elements of the set can be recovered from the reduced representation by probing with the closest item in the memory, and consecutive subtraction. Grouping is essential for defining ''a part of'', ''contains'' relationships.
  
There are two binding operations: bitwise XOR and permutation. Binding operation maps (randomizes) the vector to a completely different space, while preserving the distances between two vectors. As stated in \cite{kanerva2009hyperdimensional}, ''...when a set of points is mapped by multiplying with the same vector, the distances are maintained, it is like moving a constellation of points bodily into a different (and indifferent) part of the space while maintaining the relations (distances) between them. Such mappings could play a role in high-level cognitive functions such as analogy and the grammatical use of language where the relations between objects is more important than the objects themselves.'' 

A few representative examples to demonstrate the expressive power of hyperdimensional computing: 

\textbf{1.} We can represent pairs of objects via multiplication. $O_{A,B} = A \oplus B$ where $A$ and $B$ are two object vectors. 

\textbf{2.} A triplet is a relationship between two objects, defined by a predicate. This can similarly be formed by $T_{A,P,B} = A \oplus P \oplus B$. These types of triplet relationships are very successfully utilized for information extraction in large knowledge bases \cite{dong2014knowledge}.

\textbf{3.} A composite object can be built by binding with attribute representation and summation. For a composite object $C$,
\begin{flalign*}
  & C = X \oplus A_1 + Y \oplus A_2 + Z \oplus A_3,
\end{flalign*}  
where $A_1$, $A_2$ and $A_3$ are vectors for attributes and $X$, $Y$ and $Z$ are the values of the attributes for a specific composite object.

\textbf{4.} A value of an attribute for composite object can be substituted by multiplication. Suppose we have assignment $X \oplus A_1$, then we can substitute $A_1$ with $B_1$ by, $(X \oplus A_1) \oplus (A_1 \oplus B_1) = X \oplus B_1$. It is equivalent to saying that $A_1$ and $B_1$ are analogous. This property is essential for analogy making. 

\textbf{5.} We can define rules of inference by binding and summation operations. Suppose we have a rule stating that ''If $x$ is the mother of $y$ and $y$ is the father of $z$, then $x$ is the grandmother of $z$'' \footnote{The example is adapted from \cite{kanerva2009hyperdimensional}.}. Define atomic relationships:
\begin{flalign*}
  & M_{xy} = M_1 \oplus X + M_2 \oplus Y,\\
	& F_{yz} = F_1 \oplus Y + M_2 \oplus Z,\\
	& G_{xz} = G_1 \oplus X + G_2 \oplus Z,
\end{flalign*}
then the rule is,
\begin{flalign*}
  & R_{xyz} = G_{xz} \oplus (M_{xy} + F_{yz}).
\end{flalign*}
Given the knowledge base, ''Anna is the mother of Bill'' and ''Bill is the father of Cid'', we can infer grandmother relationship by applying the rule $R_{xyz}$:
\begin{flalign*}
  & M_{ab} = M_1 \oplus A + M_2 \oplus B,\\
	& F_{bc} = F_1 \oplus B + M_2 \oplus C,\\
	& G^{'}_{ac} = R_{xyz} \oplus (M_{ab} + F_{bc}), 
\end{flalign*}      
where vector $G^{'}_{ac}$ is expected to be very similar to $G_{ac}$ which says ''Anna is the grandmother of Cid''. Please note that the if-then rules represented by hyperdimensional computing can only be if-and-only-if logical statements because operations used to represent the rule are symmetric.

\textbf{6.} A sequence of objects can be compactly represented by convoluted permutation. Suppose we have a sequence $ABCD$, then the vector for  sequence is, 
\begin{flalign*}
  & S = \Pi( \Pi (\Pi (\Pi A) + B) + C) + D.
\end{flalign*} 
Then the stored sequence can be generated in a similar way objects in a set are recovered. 

Without losing the expressive power of classical hyperdimensional computing, we are introducing cellular automata to the framework. In our approach we will use the binary cellular automata reservoir vector as the representation of objects and predicates instead of random vectors, to be used for symbolic computation. There are two major advantages of this approach over random binary vector generation:

\textbf{1.} Reservoir vector enables connectionist pattern recognition and statistical machine learning (as demonstrated in previous sections), while random vectors can only be used for symbolic computation.

\begin{figure}
\begin{center}
\includegraphics[width=0.8\textwidth]{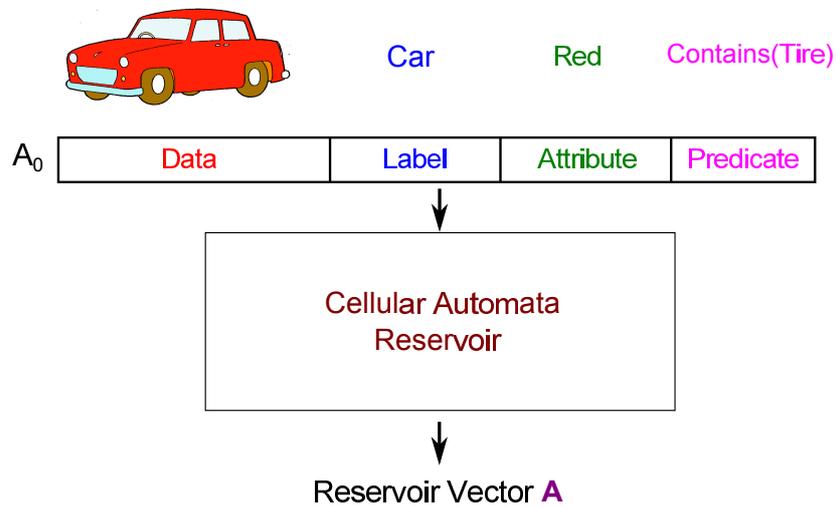} 
\end{center}
   \caption{The hyperdimensional vector generation architecture. Each object can be defined by its data i.e. image (adapted from clipartpanda.com), label, attributes and related predicates.}
\label{fig:fig7}
\end{figure}

\textbf{2.} The composition and modification of objects can be achieved in a semantically more meaningful way, by devising an architecture of cellular automata initialization. We are envisioning a structure for cellular automata initial vector (Figure \ref{fig:fig7}) that includes the  data (eg. image), object label (car), object attributes (red) and related predicates (contains). Reservoir feature is obtained by cellular automata reservoir expansion, to be used for connectionist and symbolic computation. If we want to create another reservoir vector, that uses the same data (image) but with modified label (eg. vehicle), we only change the label section of the initial vector. Then the two reservoir vectors will be similar, as they should be because they differ very little in terms of definition. The semantic similarity of the two data instances can be preserved in the reservoir hyperdimensional vector representation, and there is no straightforward mechanism for this in classical hyperdimensional computing framework. 

\begin{figure}
\begin{center}
\includegraphics[width=0.6\textwidth]{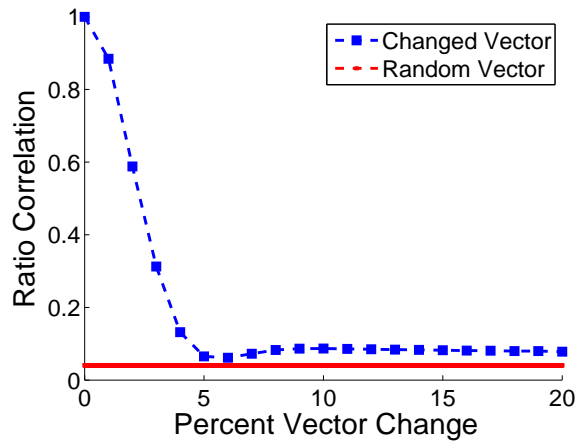} 
\end{center}
   \caption{The simulation results that investigate the effect of changing a vector on the reservoir feature. See text for details. }
\label{fig:fig8}
\end{figure}

How much difference in initial vector can be tolerated? It is related with the Lyapunov exponent of the cellular automaton, and it is demonstrated for rule 90 in Figure \ref{fig:fig8}. The ratio of correlation between original reservoir vector and the changed reservoir vector is given with respect to the percentage of change in the initial vector ($A_0$). Experiments indicate that the modified/distorted reservoir vector is inside the semantic space of the original vector, up to \% 4 change initial vector definition.  

Cellular automata reservoir vector representation for \textbf{linear CA rules} resembles MBAT architecture of Gallant et al. \cite{gallant2013representing} or sequence coding of Kanerva \cite{kanerva2009hyperdimensional}. In that approach, a sequence objects is embedded into a fixed length vector by continuous matrix multiplication (see bullet 6 of representative examples above). In addition to matrix operations, cellular automata enable a wide spectrum of nonlinear operations by using different rules. Also, our framework gives a direct way of vector computation for non-temporal tasks. Thus, it can be considered as generalization of MBAT architecture.     

\subsection{Symbolism for Linear Cellular Automata Rules}
Any cellular automaton rule can be used for reservoir expansion, and hyperdimensional computing capabilities won't change for different CA rules. However, again linear rules are giving us computational and theoretical opportunities, and they will be explored in this section. 

Linear CA rules show additive behavior: the evolution for different initial conditions can be computed independently, then the results are combined by simply adding \cite{chaudhuri1997additive, wolfram2002new}. For example, rule 90 is additive under exclusive or (XOR) operation such that, when two separate initial conditions are combined by XOR (shown by $\oplus$), their subsequent evolution can also be combined by XOR. Although the combination logic for rules 150, and 22 (because it simulates rule 90) can also be derived, we will focus on rule 90.

Suppose we have two separate inputs, $A$, $B$. Let us assume that the nonzero entries in the input (i.e. initial states of the CA) represent the existence of categorical objects, as in 5 bit/20 bit tasks. We compute the reservoir by applying the rule (i.e. 90) for a period of time steps and concatenating the state space, call them $C_A$ and $C_B$. We are interested in a new concept by combining the two inputs: $A \vee B$. This new concept should represent the union of objects, existing separately in $A$ and $B$, thus it is more abstract. Due to the binary categorical indicator nature of the input feature space, definition of logical combination rules are straightforward. We define OR operation by computing the reservoir of $A \vee B$:
\begin{align*}
    OR(A,B) = {C}_{A \vee B} = {C}_{A} \oplus {C}_{B} \oplus {C}_{A \wedge B} = {C}_{A} \oplus {C}_{B-A}.
\end{align*} 

The representation of the new concept obtained by union on existing concepts, can be computed on the cellular automata reservoir feature space via XOR operation, which is equivalent to addition operation on Galois Field, $F_2$. OR is a grouping operation that is achieved by normalized summation in classical hyperdimensional computing framework.

If the pattern is already stored in a concept, a repetitive addition does not make a change:
\begin{align*}
    {C}_{(A \vee B) \vee B} = {C}_{A \vee B} \oplus {C}_{\mathbf{0}} = {C}_{A \vee B},
\end{align*} 
and this is essential for incremental storage \cite{jaeger2014controlling}. 

AND operation will generate a concept which consists of categories that co-exist in both $A$ and $B$ \footnote{Derived using De Morgan's rule and experimentally verified.}. :
\begin{align*}
    AND(A,B) = {C}_{A \wedge B} = {C}_{A} \oplus {C}_{A-B}.
\end{align*} 

For application of AND/OR operations, initial vector of the CA reservoir needs to be saved for all logical entities in order to extract ${C}_{A-B}$ and ${C}_{B-A}$, however this is not a practical issue.  

XOR operation is straightforward: 
\begin{align*}
		XOR(A,B) = {C}_{A} \oplus {C}_{B}.
\end{align*}

In classical hyperdimensional computing, normalized summation can be used as AND in rule formation (item $5$ in section \ref{sec:CombineConnectionistSymbolic}). However the real meaning of the operation as intersection of concepts can not be utilized, thus AND does not really exist. This major advantage of CA framework is due to the semantically meaningful computation of vectors via cellular automata evolution, instead of random generation. In fact, in linear CA symbolism, it is as if operations are manipulating the initial vector that hold logical variables. Semantics of the initial vectors are preserved after every logical operation on the reservoir due to additivity, eg. when the reservoirs are OR'ed the resultant vector is identical to a reservoir computed with OR'ed initial conditions. Therefore the system is expressively equivalent to a propositional logic language. Overall it has great expressive power: availability of XOR and AND forms the whole $F_2$ field and it is possible to represent any logic obtainable by (OR , AND), with the additional benefit of algebraic operations. Yet the reservoir expansion also provides a means for statistical machine learning as demonstrated in the previous sections. The symbolic computation in linear CA rule system resembles hard conceptors in \cite{jaeger2014controlling}, although XOR operation seems missing in the latter. 

\begin{figure}
\begin{center}
\includegraphics[width=0.4\textwidth]{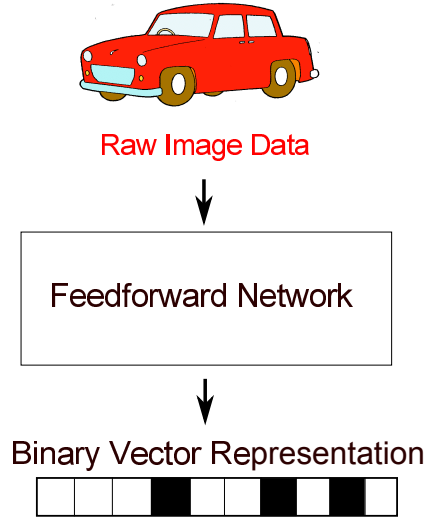} 
\end{center}
   \caption{Binary representation for data that is inherently non-binary (for example an image) can be obtained by binarizing the feedforward network representation of the data.}
\label{fig:fig9}
\end{figure}

\subsection{Experiments on Cellular Automata Symbolic Computation}
\label{sec:BinarizeData}
It should be noted that, the nice symbolic computation properties of the linear cellular automaton framework is applicable when the non-zero feature attributes of the CA initial states represent the existence of a predefined object/concept \footnote{Binary coding for initial conditions of the CA is another option but logical operations are meaningless in this case.}. This is not an issue for object labels, attributes or predicates (Figure \ref{fig:fig7}), but sensory data is not necessarily categorical. Having said that, any sensory data space can be transformed into categorical indicator space by using a feedforward neural network in the front-end (Figure \ref{fig:fig9}) \footnote{Another option for categorical transformation is quantization of the data (or the frequency spectrum (DFT or Hadamard) of the data) and then encoding the quanta in binary, which might not be practical for high dynamic range problems. }. This approach is parallel to the three stage proposal by Gallant et al. \cite{gallant2013representing}, in which preprocessing stage provides the required feature expansion.  

We have previously shown that binarization of the hidden layer activities of a feedforward network is not very detrimental for classification purposes (\cite{yilmaz2015HolisticVision} under review). For an image, the binary representation of the hidden layer activities holds an indicator for the existence of Gabor like corner and edge features, hence provides the means for the required categorical indicator space. The proposed perspective on feedforward networks generalizes the Local Binary Pattern (LBP) analysis \cite{ojala1994performance}. This capability can be generalized to domains other than images, and higher level descriptions of objects. 

In order to test the performance of CA features on binarized hidden layer activities, we run experiments on CIFAR 10 dataset. We used the first 500 training/test images and obtained single layer hidden neuron representation using the algorithm in \cite{coates2011analysis} (200 number of different receptive fields, receptive fields size of 6 pixels). The neural activities are binarized according to a threshold and on average 22 percent of the neurons fired with the selected threshold. After binarization of neural activities, CA features can be computed (rule 90, $R=4$ and $I=4$) on the binary representation as explained in section \ref{sec:method_section}. We trained linear and RBF kernel SVM classifiers \cite{CC01a} on both real and binary neural activities, as well as CA features that are computed on the binary neuron activities. The parameters of the SVM classifiers are optimized using a coarse-to-fine grid search, and CA feature experiment is repeated many times due to randomized permutation. The results are given in Table \ref{table:CIFAR10CA}. It is observed that binary neuron shows superior performance over real neuron, and this needs to be investigated further. But more importantly, CA feature on linear classifier exceeds (best random permutation) the RBF kernel nonlinear classifier. Yet, the computational complexity is much lower (the expansion is $R \times I = 16$) and it has a potential for improvement via hybridization and multilayer architecture (see Discussion).   


\begin{table}[h]                 
\centering                        
\begin{tabular}{l|l|l|l|l|l} 
  & Linear & RBF \\  
\hline  
Real Neuron & 34.6 & 35.2  \\       
Binary Neuron & 37.6 & 39.6 \\
CA Feature (max/avg) & 41.8/39 & -   
\end{tabular}                            
\caption{The experiments on CIFAR 10 dataset (subset). The classification performance for real and binary neural representation is given for linear and RBF kernels. The CA feature is computed on binary neuron activities, and linear kernel performance on CA feature (max and average are given) is equivalent to RBF kernel performance on binary neurons. }          
\label{table:CIFAR10CA}        
\end{table} 

We also tested the classification performance of CA conceptors as in \cite{jaeger2014controlling}. We formed a separate conceptor for each class using binary neural representation (rule 110, $R=16$ and $I=16$, 50 samples for eac class) of CIFAR training data and vector addition defined in \cite{snaider2012integer}. Then we tested the classification performance of the conceptors on test data, such that each test data CA feature vector is associated with the closest (max inner product) conceptor. The results are given in \ref{table:ConceptorCA}. It is observed that classification with CA conceptor outperforms the classification with hidden layer conceptor, which supports the idea that CA expansion enhances the symbolic computation. More importantly, conceptor classification provides a much more flexible framework discussed in \cite{jaeger2014controlling}, where addition/deletion of data and classes is straightforward and logical operation are enabled on the semantic space.  

\begin{table}[h]                 
\centering                        
\begin{tabular}{l|l} 
Hidden Layer Conceptor  & CA Conceptor \\  
\hline  
28.3 & 32.5  \\       
\end{tabular}                            
\caption{The experiments on CIFAR 10 dataset (subset). The classification performance of conceptors that are computed from binary hidden layer activities and cellular automata features for $R=16$ and $I=16$.}          
\label{table:ConceptorCA}        
\end{table} 

In order to demonstrate the power of enabled logical operation, we will use analogy. Analogy making is crucial for generalization of what is already learned. We tested the capability of our symbolic system using images. The example given here follows "What is the Dollar of Mexico?" in \cite{kanerva2009hyperdimensional}. However in the original example, sensory data (i.e. image) is not used, because there is no straightforward way to introduce sensory data into hyperdimensional computing framework. The benefit of using non-random binary vectors is obvious in this context. 

We formed two new concepts called \textbf{Land} and \textbf{Air}:
\begin{align*}
		Land = Animal \oplus Horse + Vehicle \oplus Automobile, \\
		Air = Animal \oplus Bird + Vehicle \oplus Airplane. 
\end{align*}
In these two concepts, CA features of Horse and Bird images are used to bind with the Animal filler, CA features of Automobile and Airplane images are used to bind with the Vehicle filler \footnote{50 training images for each, rule 110, $R$ and $I$ are both 16.}. Animal and Vehicle fields are represented by two random vectors \footnote{22 percent non-zero elements}, the same size as the CA features. Multiplication is performed by xor ($\oplus$) operation and vector summation is again identical to \cite{snaider2012integer}. The final products, Land and Air are also CA feature vectors, and they represent the merged concept of observed animals and vehicles in Land and Air respectively. We can ask the analogical question \textbf{"What is the Automobile of Air?"}, $AoA$ in short. The answer is simply given by:
\begin{align*}
		 AoA = Automobile \oplus Land \oplus Air.  
\end{align*}
$AoA$ is a CA feature vector, and expected to be very similar to Airplane conceptor. We tested the analogical accuracy using unseen Automobile test images (50 of them), computing their CA feature vectors followed by $AoA$ inference, then finding the closest conceptor class to $AoA$ vector (max inner product). It is expected to be the Airplane class. The results of this experiment is given in Table \ref{table:AnalogyCA}. The analogy on CA features is 67 percent accurate, whereas if the binary hidden layer activity is used instead of CA features (corresponds to $R$ and $I$ equal to 1) analogy is only 21 percent accurate. This result clearly demonstrates the benefit of CA feature expansion on symbolic computation, even more than the classification task given in Table \ref{table:ConceptorCA}. 

\begin{table}[h]                 
\centering                        
\begin{tabular}{l|l} 
Hidden Layer Feature & CA Feature\\  
\hline  
21.3 & 67.2  \\       
\end{tabular}                            
\caption{The analogy making experiment on CIFAR 10 dataset (subset). The accuracy of the analogy is given for binary hidden layer neuron representation and CA features. See text for details. }          
\label{table:AnalogyCA}        
\end{table} 

The analogy given above implicitly assumes that, Automobile concept is already encoded in the concept of Land. What if we ask "What is the Truck of Air?" ? Even though Truck images are not used in building the Land concept, due to the similarity of Truck and Automobile concepts we might still get good analogies. The results on these second order analogies are contrasted in Table \ref{table:AnalogyCA2}. Automobile and Horse (i.e. "What is the Horse of Air?", the answer should be Bird.) are first order analogies and they give comparably superior performance as expected, but second order analogies are much higher than chance level (10 percent). 

\begin{table}[h]                 
\centering                        
\begin{tabular}{l|l|l|l|l|l} 
\textbf{Automobile} & \textbf{Horse} & Truck & Deer & Ship & Frog \\  
\hline  
67.2 & 58.2 & 38.6 & 38.9 & 31.5 & 24.4 \\       
\end{tabular}                            
\caption{Another analogy making experiment on CIFAR 10 dataset (subset). The accuracy of the analogy is given for first (given in bold) and second order analogies. See text for details. }          
\label{table:AnalogyCA2}        
\end{table} 

Please note that, these analogies are performed strictly on the sensory data, i.e. images. Given an image, the system is able to retrieve a set of relevant images that is linked through a logical statement. A very small number of training data is used, yet we can infer conceptual relationships between images surprisingly accurately. And it is possible to build much more complicated concepts using hierarchies, for example Land and Air are types of environments and can be used as fillers in Environment field. Ontologies are helpful to narrow down the set of required concepts for attaining a satisfactory description of the world. Other modalities such as text data are also of great interest, (see \cite{mikolov2013distributed} for state-of-the-art), as well as information fusion on multiple modalities (eg. image and text).   

\begin{figure*}
\begin{center}
\includegraphics[width=1\textwidth]{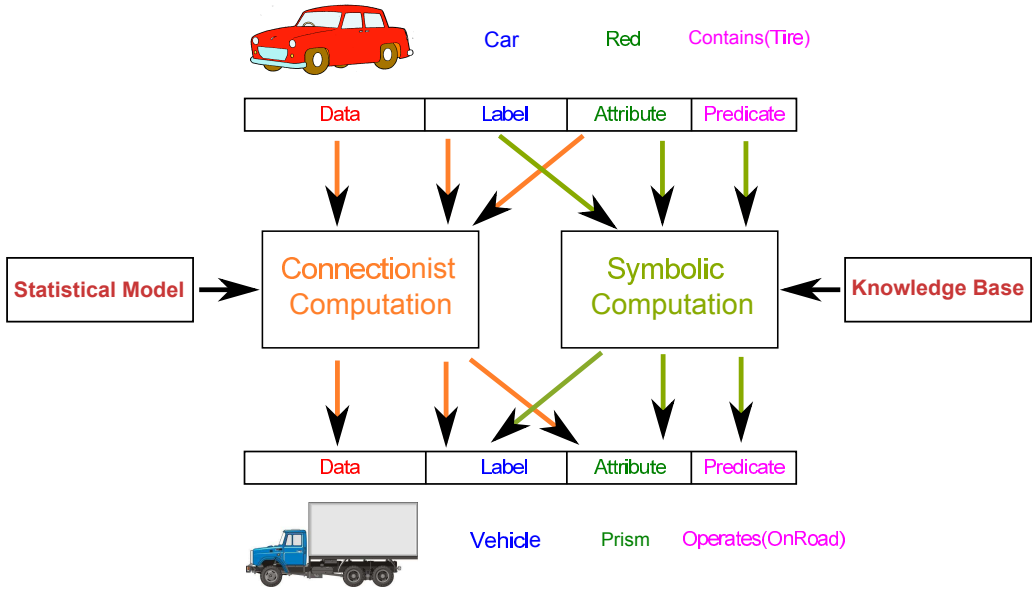} 
\end{center}
   \caption{Combining connectionist and symbolic processing through reservoir and hyperdimensional computing. See text for details.}
\label{fig:fig10}
\end{figure*}

\section{General Architecture for Connectionist-Symbolic Machine Intelligence}
Reservoir computing and hyperdimensional computing can work together to achieve simultaneous connectionist and symbolic intelligence, as shown in Figure \ref{fig:fig10}. 

Through reservoir computing based machine learning it is possible to achieve these partial estimations:\\
\textbf{1.} Estimate object label when data is given (classification).  \\
\textbf{2.} Estimate object attributes when data is given (regression).  \\
\textbf{3.} Estimate object label when attributes are given (descriptive models).  \\
\textbf{4.} Generate data when attributes and label is given (generative statistical model).

More importantly, when these partial estimations are performed, the initial sensory processing is done and memory retrieval is enabled. The $data+label+attribute$ vector is projected on the CA reservoir space and it is stored as Sparse Distributed Memory (SDM) \cite{kanerva1993sparse}. Then SDM based retrieval works in tandem with machine learning based partial estimations as an autoassociative memory, to fill-in missing data and improve precision/accuracy (see also \cite{yilmazFastRecurrent} for a low complexity option). We conjecture that this retrieval is the essence of recurrent computation in human cortex. 

Using the power of hyperdimensional computing we can use labels, attributes and predicates to achieve:\\
\textbf{1.} Composite object description via attributes and predicates.  \\
\textbf{2.} Analogy making.  \\
\textbf{3.} Rule based inference. \\

These are partial symbolic estimations. After performing the possible logical inferences, $label+attribute+predicate$ vector is projected on the CA reservoir space and again stored as SDM. Similarly, SDM based retrieval is enabled that fills-in and regularizes incomplete knowledge. 

As the data structure is filled in with extracted knowledge through simultaneous processing of the two streams, such as more detailed attributes for an object (color, shape, material, size, symmetry), a class label hierarchy (vehicle-car-sedan), or an intricate relationship of predicates (Contains(Car,Tire), isPartOf(Tire,Vehicle), isBelow(Tire,Window)), the vector becomes a very rich description of the object. It should be mentioned that both the statistical model and knowledge base are dynamic entities in this picture: they should adapt and evolve to include the learned knowledge. 

Data, labels and attributes are projected onto reservoir feature space for sensory processing; labels, attributes and predicates are projected onto reservoir feature space for non-traditional (i.e. hyperdimensional) symbolic processing. Thus, both low level data and high level knowledge live on the same space. After partial estimations and retrievals, we have a more complete $data+label+attribute+predicate$ vector. This is a blackboard of a cognitive session. This whole vector can be projected onto the CA reservoir feature space, and saved as a semantic memory \cite{squire1992declarative}. 

In the proposed architecture, there is a \textbf{cross-talk between connectionist and symbolic computation} via labels and attributes. The framework offers a potential for holistic artificial intelligence, where we can use the results of pattern recognition on logical inference and vice versa. 

We can give a few examples for possible usage scenarios:\\
1. Object-scene relationship in an image can be handled using a high order language, where we can define complicated logical rules: if we detect a computer object and a nearby desk object then we infer that it is an office scene unless, there isn't a car, tree or building objects in the image. \\
2. We can infer hidden states: in an image if we detect a face object, but can not detect a human object, and also we detect a chair object nearby the face object, then the person is sitting. \\
3. We can start a search for possible objects: if we classify the scene as street, start a search for cars and humans. \\
4. We can detect parts of an object: if we detect a car in the scene, through Contains predicate we can search for its parts such as tires, windows and lights. If we can not detect its parts, we can raise a flag via Occluded predicate. We can infer its pose via detected parts: if two tires, two windows are detected but no lights are detected the pose is sideways. If we infer the pose, then we can repeat the detection and search with this knowledge, possibly with a more refined classifier. \\
5. Sensory feedforward flow forms beliefs about the external world while goal based feedback can cause priors, and this is mainly achieved by SDM retrievals. We can use logical statements to define goals and these can modify labels and attributes, hence gate sensory perception.     

Simultaneous connectionist and symbolic computation is a grand claim. It not only requires a complete system where domain dependent ontologies, procedural memory and production execution is included \cite{anderson2014atomic}, but also deserves a thorough examination, which is planned as a future work.

\section{Computational Complexity}
There are two major savings of cellular automata framework compared to classical echo state networks:

1. Cellular automaton evolution is governed by bitwise operations instead of floating point multiplications. 

2. Since the reservoir feature vector is binary, matrix multiplication needed in the linear classification/regression can be replaced by summation. 

Multiplication in echo state network is replaced with bitwise logic (eg. XOR for rule 90) and multiplication in classification/regression is replaced with summation. Overall, multiplication is completely avoided, both in reservoir computation and in classifier/regression stages, which makes the CA framework especially suitable for FPGA hardware implementations. 

\begin{table}[h]                 
\centering                        
\begin{tabular}{l|l|l|l} 
Task & ESN (Floating Point) & CA (Bitwise) & Speedup  \\  
\hline     
5 bit $T_0=200$ & 1.03 M & 0.43 M & 2.4X  \\
5 bit $T_0=1000$ & 13.1 M & 8.3 M  & 1.6X \\  
20 bit $T_0=200$ & 17.3 M & 9.5 M  & 1.8X \\   
\end{tabular}                     
\caption{The comparison of the number of operations for the echo state networks (ESN) and the Cellular Automata (CA) framework. Operation is floating point for ESN, but bitwise for CA.}
\label{table:ComparisonOperation}        
\end{table}

The number of operations needed for the reservoir computation of 5 bit and 20 bit tasks is given in Table \ref{table:ComparisonOperation}, both for Echo State Network (ESN) in \cite{jaeger2012long} and for cellular automata (CA). \footnote{For ESN, it is assumed that the number of floating point operations is equal to 2*NNZ.} There is a speedup in the number of operations in the order of 1.5-3X. However, considering the difference of complexity between floating point and bitwise operations, there is \textbf{almost two orders} of magnitude speedup/energy savings. \footnote{CPU architectures are optimized for arithmetic operations: bitwise logic takes 1 cycle and 32 bit floating point multiplication takes only 4 cycles, on 4th generation Haswell Intel\textsuperscript{TM} core. Therefore the speedup/energy savings due to the bitwise operations will be much more visible on hardware design, i.e. FPGA.}

When support vector machines are used for linear cellular automaton rule kernels, the computational complexity is $\mathcal{O}(N)$, in which $N$ is the size of the initial data vector. However computational complexity is $\mathcal{O}(NIR)$ for general rules, where $R$ is the number of permutations and $I$ is the number of CA evolutions \footnote{Because linear methods are applied on reservoir feature space of size $NIR$.}. Given the fact that interesting problems would require $R>30 \text{ and } I>30$ (eg. Table \ref{table:20BitGoL200and300}), then usage of linear CA rule kernels will provide at least \textbf{three orders of magnitude} speedups \footnote{For 20 bit task, $R \times I \approx 6000$.}. The same argument is valid when we compare CA kernel with nonlinear kernels (eg. RBF), where the complexity is $\mathcal{O}(DN)$, where $D$ is the number of support vectors in the order of thousands for interesting problems. 

For symbolic processing, there is \textbf{no additional computation} for CA framework, reservoir outputs can directly be combined using logical rules as in hyperdimensional computing. However, Conceptors \cite{jaeger2014controlling} that are built upon ESN require correlation matrix computation and matrix multiplication of large matrices, for each input. As an example, for 20 bit task $T_0=200$, 1760 M floating point operations are needed for correlation matrix computation. Then there is a matrix inversion ($2000 \times 2000$ size, 68 M operations) and matrix multiplication (two $2000 \times 2000$ size matrices, 16000 M operations) to obtain the Conceptor matrix. All these computations (about 18 billion) are avoided in our framework.

\section{Discussion}
\label{discussion_section}
We provide a novel feedforward and reservoir computing frameworks that are capable of long-short-term-memory and symbolic processing, which requires significantly less computation compared to echo state networks (Table \ref{table:ComparisonOperation}  ). \footnote{Deatiled comparison with other RNN algorithms are not provided but the computational complexity argument seems to be generally valid.} In the proposed approach, data are passed on a cellular automaton instead of an echo state network (Figure \ref{fig:figm1}), and similar to echo state networks with sparse connections, the computation is local (only two neighbors in 1D, implying extreme locality) in the cellular automata space. Several theoretical advantages of the cellular automata framework compared to echo state networks are mentioned, in addition to their practical benefits. Cellular automata are easier to analyze, have insurances on Turing completeness and allows Boolean logic as well as algebra on Galois Field. From CA side of the medallion: the computation performed in cellular automata can be conceptualized in many levels \cite{mitchell1996computation,jamesevolutionary}, but \textbf{our main proposition is that, reservoir computing is a very good fit for harnessing the potential of cellular automata computation.}

\subsection{Distributed Representation on CA states}      
Some of the best performing cellular automata rules are additive. How does extremely local, additive and bitwise computation gives surprisingly good performance in a pathological machine learning task? This question needs further examination, however the experiments (see Section \ref{sec:results_section}) suggest that if a dynamical system has universal computation capability, it can be utilized for difficult tasks that require recurrent processing once it is properly used. The trick that worked in the proposed framework is multiple random projections of the inputs that enhanced the probability of long range interactions (Table \ref{table:5BitElem} ). 

In the paper we prove that cellular automata feature space holds a distributed representation of higher order attribute statistics, and it is more effective than local representation (Table \ref{table:5BitElemCA_LocalRep}). Second order statistics (covariance matrix) is shown to be very informative for a set of AI tasks \cite{tuzel2006region}, yet we are providing a novel way of exploiting higher order statistics using cellular automata. 

The usage of cellular automata for memorizing attribute statistics is in line with Elementary Callular Automata with Memory (ECAM). ECAM uses history by including previous states in every cell iteration. Reservoir computing based CA proposed in this paper uses classical CA rules, do not crash current cell states with previous states, but record all the history in a large reservoir. \cite{martinez2013designing,alonso2006elementary} show that the memory function can push a CA rule into a completely different computation regime and this can be used in our framework for rule hybridization process in the reservoir. Also ECAM will provide a neat way of handling sequences without the need for flattening. 

\subsection{Cellular Automata Kernel Machines}
As it is shown in \cite{tuzel2008pedestrian} that the distance metrics in covariance statistics can be learned, we provide the theoretical foundation for distance metric learning of CA features for linear cellular automata rules (section \ref{sec:metriclearning_section}). We show that the distance estimates are very accurate for a wide range of parameter settings (Figure \ref{fig:fig4}). More importantly, linear cellular automata features can be kernelized to be used in support vector machine framework (section \ref{sec:kernellearning_section}), and they have the same computational complexity with linear kernel while approaching the performance of nonlinear kernels (Table \ref{table:MNIST_KernelTwoFeatures}). Linear CA rules are shown to be powerful computational tools \cite{baetens2010phenomenological,martinez2013note,alonso2006elementary,zenil2013asymptotic}, yet we can use very low complexity kernel methods to estimate their feature space. The kernelization effort is parallel to recurrent kernel machines framework \cite{hermans2012recurrent,shi2007support} and further experiments are needed to quantify the performance of CA kernelization and comparison with recurrent kernel machines. 

\subsection{Symbolic Computation and Beyond}
Along with the pattern recognition capabilities of cellular automata based reservoir computing, hyperdimensional computing framework enables symbolic processing. Due to the binary categorical indicator nature of the representation, rules that make up the knowledge base and feature representation of data that make up the statistical model live on the same space, which is essential for combining connectionist and symbolic capabilities. It is possible to make analogies, form hierarchies of concepts and apply logical rules on the reservoir feature vectors.

In the experiments (Tables \ref{table:CIFAR10CA} and \ref{table:ConceptorCA}), we showed that CA based symbolic computation approaches the performance of statistical machine learning based classification, i.e. SVM. Yet, we haven't searched for any type of optimization, or even tried the symbolic system with very large vectors, so there is room for improvement. Additionally, conceptor based classification provides a much more flexible framework for data and class maintenance, with the added benefit of logical operations/queries. To illustrate the logical query, we have shown the capability of the system to make analogies on image data. We asked the question "What is the Automobile of Air?" after building Land and Air concepts out of images of Horse, Automobile (Land), Bird and Airplane (Air). The answer correct is Airplane and the system infers this relationship with 67 percent accuracy. 

Another future direction for the potential application of the framework is data fusion. The proposed data representation (Figure \ref{fig:fig7}) can be used for fusion on multiple levels: sensor level using data and label; knowledge level using attribute and predicates. We can concatenate multiple objects one after another, such as an image object and a text (i.e. image caption) object. Then the reservoir will compute inter-object statistics, capturing cross-modal interactions. Although there is exciting advancement in simultaneous image-text learning in neural network literature \cite{kiros2014unifying}, they are strictly in the realm of statistical machine learning being incapable of representing a high level knowledge bases and logical rules.       

\subsection{Extensions and Implementation}
There are a few extensions of the framework that is expected to improve the performance and computation time:\\
1. A hybrid \cite{sipper1998evolving} and a multilayer automaton can be used to handle different spatio-temporal scales in the input. Two types of hybrid, in one is CA space is hybrid, in another after each random permutation a different rule is applied.  \\
2. The best rule/random mapping combination can be searched in an unsupervised manner (pre-training). The rank of the binary state space can be used as the performance of combinations. Genetic algorithm based search is also a good option for non-elementary rules \cite{das1994genetic}.\\
3. We can devise ECAM rules in the reservoir instead of classical rules. In that approach we will have another way of achieving long range attribute interactions, and hybridization. 

Cellular automata are very easy to implement in parallel hardware such as FPGA (\cite{halbach2004implementing}) or GPU (unpublished experiments on \cite{MarkFiserWeb}). 100 billion cell operations per second seem feasible on mid-range GPU cards, this is a very large number considering 10 million operations are needed for 20 bit task.

\subsection{Recurrent Computation in Cellular Automata Reservoir}
We provide a recurrent architecture (section \ref{sec:CA_Reservoir_Recurrent}), that holds a fixed sized representation (cellular automata state vector) for a sequence of arbitrary length. The proposed algorithm is able to generate data as in echo state networks. This capability needs to be tested, possibly using language task such as character and word prediction. The results on noiseless memorization task is encouraging, yet there are many possibilities for improvement, some of them mentioned above.  

\subsection{Potential Problems}
There is one superficial problem with the proposed framework: reservoir expansion is expected to vastly increase the feature dimension for complicated tasks ($R$ and $I$ are both very large), and curse with dimensionality. However, linear kernel that will be used in large reservoir feature space is known to be very robust against large dimensionality (deduced using structural risk minimization theory, \cite{vapnik1998statistical}). Although linear kernel behaves nicely, the remaining problem due to the large dimensionality of the feature space can be alleviated by using a bagging approach that also selects a subset of the feature space in each bag \cite{latinne2000mixing} or the kernel method proposed in section \ref{sec:kernellearning_section} for linear CA rules. 

As a future work we would like to test the framework on large datasets for language modeling, music/handwriting prediction and computer vision. Symbolic processing performance of the proposed framework needs to be evaluated in detail. 

\section{Acknowledgments}
This research is supported by The Scientific and Technological Research Council of Turkey (TUB\.{I}TAK) Career Grant, No: 114E554 and Turgut Ozal University BAP Grant, No: 006-10-2013. 

\newpage
\appendix
\section{Details of Experiments} \label{App:Appendix}
Here we provide the details of the experiments performed in the paper. The Matlab\textsuperscript{\textregistered}codes for all experiments are shared in \href{http://www.ozguryilmazresearch.net}{ozguryilmazresearch.net}. The appendix is partitioned into sections the same name as the ones in the main body of the paper. 

\subsection{Cellular Automata Reservoir}
We have a binary data vector of size $N$. We permute the data vector, compute cellular automata states for a fixed period of time and concatenate to get CA reservoir feature vector. 
The computation is done in a nested for loop:\\
for permutation = 1 to $R$ \\
\indent Retrieve a random permutation \\
\indent	Permute data \\
\indent	for iteration = 1 to $I$ \\
\indent \indent			Compute next state of CA \\
\indent \indent			Store in a matrix \\
\indent  end for loop \\
\indent	Store CA evolution in a matrix \\
end \\
Concatenate the CA evolution matrix, to get a vector of size $N \times I \times R$. The only difference in Game of Life is that, we map the input data vector onto a 2D square grid of suitable size during permutation.

\subsection{Memory of Cellular Automata State Space}
In these experiments, the Matlab\textsuperscript{\textregistered} code kindly provided by \cite{jaeger2012long} is used without much change. The echo state network based reservoir computation is replaced by cellular automata reservoir. The sequence is flattened by concatenating the time steps into a large binary vector. Then the CA reservoir receives it as the initial conditions, and computes the evolution for many different random permutations. Inclusion of the original data vector to the CA feature vector does not alter the performance significantly (interestingly exclusion performs better), and it is included. Increasing the distractor period expands the size of the vector, and adds irrelevant data to the CA computation reducing the estimation quality. Pseudoinverse based regression is used to estimate the output vector, all at once. 300 data points are used for training in 20 bit task, and only 10 time steps are used from the distractor period (called NpickTrain) during training. Yet, regression is expected to give the vector for full distractor period. 100 test data are used for test. 25 trials are averaged.    

\subsection{Distributed Representation of Higher Order Statistics}
In this experiment we used the 5 bit memory task to compare the distributed and local representation of attribute statistics. For doing that, we computed the ${C}_{k}$ vectors instead of ${A}_{k}$ (CA evolution at $k_{th}$ time step) for each time step and each permutation and formed an identical sized feature vector:
\begin{align*}
	C_1 = A_1 \\
	C_2 = A_2 \\
	C_3 = A_3 \oplus A_1 \\
	C_4 = A_4 \\
	C_5 = A_5 \oplus A_3 \oplus A_1 \\
	C_6 = A_6 \oplus A_1 \\
	C_7 = A_7 \oplus A_5 \oplus A_1 \\
	C_8 = A_8 \\
	FeatureVector = [C_1; C_2; C_3; C_4; C_5; C_6; C_7; C_1].
\end{align*}

Then we used that vector for regression.

\subsection{Metric Learning for Linear Cellular Automata Reservoirs}
The information given in the main body is enough to replicate the simulations. 

\subsection{Linear Complexity Kernel for Support Vector Machine Learning}
The information given in the main body is enough to replicate the experiment performed on MNIST dataset. 

\subsection{Computational Power}
The information given in the main body is enough to replicate the simulations. 

\subsection{Cellular Automata Reservoir}
In these experiments, again the Matlab\textsuperscript{\textregistered} code kindly provided by \cite{jaeger2012long} is used without much change. The echo state network based reservoir computation is replaced by cellular automata reservoir. The details of the reservoir features are given in section \ref{sec:CA_Reservoir_Recurrent}. Pseudoinverse based regression is used to estimate each sequence output, one at a time. 20 trials are averaged. 

\subsection{Experiments on Cellular Automata Symbolic Computation}
In order to test the performance of CA features on binarized hidden layer activities, we run experiments on CIFAR 10 dataset. We used the first 500 training/test images. We obtained single layer hidden neuron representation using the algorithm in \cite{coates2011analysis}. The code is provided by Coates on the web, \href{http://www.cs.stanford.edu/~acoates/}{here}. First, 200 number of different receptive fields of size 6 pixels are learned using k-means on a million randomly cropped image patches. Then each 6 by 6 image region is sparsely encoded by the similarity to 200 receptive fields. The similarity measure is a real number, but we need a binary representation in order to run CA rules. The neural activities are binarized according to a threshold and on average 22 percent of the neurons fired with the selected threshold. After binarization of neural activities, CA features can be computed (rule 90, $R=4$ and $I=4$) on the binary representation as explained in section \ref{sec:method_section} and detailed in the appendix above. We trained linear and RBF kernel SVM classifiers \cite{CC01a} on both real and binary neural activities, as well as CA features that are computed on the binary neuron activities. The parameters of the SVM classifiers are optimized using a coarse-to-fine grid search, and CA feature experiment is repeated 20 times due to randomized permutation.

We also tested the classification performance of CA conceptors as in \cite{jaeger2014controlling}. We formed a separate conceptor for each class using the CA feature vectors, derived from binary neural representation explained above. Rule 110 is run for $R=16$ and $I=16$ and a CA feature vector is obtained for each image and 50 samples are used for each class of CIFAR training data. Denote $F_{ij}$ for $j_{th}$ sample of $i_{th}$ class. 

Vector addition defined in \cite{snaider2012integer} is used to form the conceptors of each class:
\begin{align*}
	\displaystyle \mathbf{T_i} = sign( \sum_{j=1}^{50} \frac{F_{ij}-0.5} {0.5} ).
\end{align*}

Then we tested the classification performance of the conceptors on test data, such that each test data CA feature vector ($F_{j}$) is associated with the closest conceptor using max inner product with the conceptor:
\begin{align*}
	c_j = \underset{i} {\mathrm{argmax}} ~ F_{j} {T_i},
\end{align*}
where $c_j$ is the class decision on the $j_{th}$ test data instance. 
The runs are repeated for 20 trials due to randomness at CA permutation, and average accuracy is reported. 

In the analogy experiments Animal and Vehicle fields are randomly created binary vectors (22 percent non-zero elements). These two vectors are used to form the Land and Air concepts using the equation in the corresponding section. 

\newpage
\section*{References}

\bibliography{egbib_oy}

\end{document}